\documentstyle [12pt]{article}
\newcommand{\beq}{\begin{equation}}
\newcommand{\enq}{\end{equation}}
\newcommand{\und}{\underline}
\newcommand{\biz}{\begin{itemize}}
\newcommand{\eiz}{\end{itemize}}
\newcommand{\di}{\displaystyle}
\newcommand{\bc}{\begin{center}}
\newcommand{\ec}{\end{center}}

\newcommand{\pa}{\partial}
\newcommand{\g}{\gamma}
\newcommand{\ep}{\epsilon}
\newcommand{\cp}{\stackrel{\circ}{p}}
\newcommand{\ck}{\stackrel{\circ}{k}}

\newcommand{\uz}{\stackrel{\circ}{U}}
\newcommand{\az}{\stackrel{\circ}{A}}
\newcommand{\bz}{\stackrel{\circ}{B}}
\newcommand{\vz}{\stackrel{\circ}{V}}

\newcommand{\nin}{\noindent}
\newcommand{\vv}{\vskip 0.25 cm}

\begin{document}
{\sf 
\vskip 2 true cm

\centerline{\bf JET-LIKE QED PROCESSES : }
\vskip 0.5 true cm 
\centerline{\bf ON GENERAL PROPERTIES OF IMPACT FACTORS}

\vskip 2 true cm
\centerline{{\bf Christian CARIMALO}\footnote{{\bf christian.carimalo@upmc.fr}}}


\vskip 2cm
\centerline{ABSTRACT}

\vv
\vv \nin In this article, general properties of impact factors involved in jet-like QED processes are explored. In particular, a general link is established between their helicity properties and their order of magnitude as defined by jet-like kinematics. Exact results are given for some processes in the strict forward direction and a general method is proposed to track orders in the calculation of multi-bremsstrahlung processes.

\newpage

\section{Preliminaries}

\vv \nin Let us first remind the way one defines an helicity frame for each particle taking part in a given process (see also Ref {\bf[1]}).

\vv \nin The 4-vectors of the basis defining the lab frame will be denoted by
$T,~X,~Y,~Z$. In this basis, the 4-momentum $p$ of a particle of mass
$m$ is developed as

\beq
p = E ~T + P \cos \theta ~Z + P \sin\theta \left( \cos\varphi~X +
\sin \varphi~Y\right)
\enq

\vv \nin where $P = \sqrt{E^2 -m^2}$. We first introduce an helicity triad $X',~Y',~Z'$
of space-like 4-vectors associated with $T$, which achieves an helicity
coupling scheme between the two time-like 4-vectors $T$ and $p$ :

\begin{eqnarray}
&X'= -\sin \theta ~Z+ \cos\theta \left( \cos\varphi~X + \sin \varphi~Y\right),~~~
Y'= -\sin\varphi~X + \cos \varphi~Y& \nonumber \\
&Z'= \cos \theta ~Z + \sin\theta \left( \cos\varphi~X +
\sin \varphi~Y\right)&
\label{hframe1} \end{eqnarray}

\vv \nin The helicity frame associated with the
4-momentum $p$ is then defined by the following 4-vectors

\beq
t = \di{p\over m},~~~x= X',~~~y=Y',~~~z= \sinh\chi~T + \cosh\chi~Z'
\label{hframe2} \enq

\vv \nin where $ \cosh\chi = E/m$.

\vv \nin The Lorentz transformation ${\cal S}$ that tranforms the reference basis
$T,~X,~Y,~Z$
into the helicity basis $t,~x,~y,~z$ is the product of three transformations :
a boost along the $Z-$axis with rapidity $\chi$, followed by a rotation
of angle $\theta$ around the $Y-$axis, followed by a rotation of angle $\varphi$
around the $Z-$axis.

\vv \nin Dirac spinors associated with the reference frame may be defined as follows.
In the spinorial representation, spin operators associated with the
reference basis $T,~X,~Y,~Z$ are
\beq
S_X = \di{1\over 2}\g_5 \g(X) \g(T),~~~S_Y = \di{1\over 2}\g_5 \g(Y) \g(T),~~~
S_Z = \di{1\over 2}\g_5 \g(Z) \g(T)
\enq

\vv \nin where the notation $\g(V) = V_\mu \gamma^\mu$ is used. 
The reference Dirac spinors will be denoted by $U^{\lambda}_0$ ($\lambda
= \pm 1/2$). They
satisfy the relations
\beq
\g(T) U^{\lambda}_0 = U^{\lambda}_0,~~~S_z U^{\lambda}_0 =
\lambda  U^{\lambda}_0,~~~{\bar U}^{\lambda}_0 U^{\lambda'}_0 = 2 \,
\delta_{\lambda' \lambda}
\enq

\vv \nin The Dirac spinors $U^{\lambda}$ associated with the helicity basis
$t,~x,~y,~z$ are related
to the reference Dirac spinors through the Lorentz transformation
\beq
{\cal S} = {\cal R}_Z(\varphi) {\cal R}_Y(\theta) {\cal H}_Z(\chi)
\label{TR1}
\enq

\nin where, in spinorial representation,

\begin{eqnarray}
&{\cal R}_Z(\varphi) = \cos(\di{\varphi\over 2}) - 2 i \sin(\di{\varphi\over 2}) S_Z&
\label{TR2}  \nonumber \\
&{\cal R}_Y(\theta) = \cos(\di{\theta\over 2}) - 2 i \sin(\di{\theta\over 2})
S_Y  &  \\
&{\cal H}_Z(\chi) = \cosh(\di{\chi\over 2}) + 2
\sinh(\di{\chi\over 2}) \g_5 S_Z & \nonumber
\end{eqnarray}

\vv \nin Explicitly, we have (using the notation $\uparrow$ for $\lambda = + 1/2$ and
$\downarrow$ for $\lambda = + 1/2$, defining spinors $V = \g_5 U$ and here
choosing the normalisation ${\bar U}^{\lambda} U^{\lambda'} = 2 m \,
\delta_{\lambda' \lambda}$)
\begin{eqnarray}
&U^{\uparrow} = \cosh(\di{\chi\over 2})U^{'\uparrow} +
\sinh(\di{\chi\over 2})V^{'\uparrow}& \nonumber \\
\label{spinor1}
&{\rm where}~~~U^{'\uparrow}= \sqrt{m} \left[\exp(-i\di{\varphi\over 2})
\cos(\di{\theta\over 2})U^{\uparrow}_0 + \exp(i\di{\varphi\over 2})
\sin(\di{\theta\over 2})U^{\downarrow}_0 \right]&  \label{sp1} \\
&U^{\downarrow} = \cosh(\di{\chi\over 2})U^{'\downarrow} -
\sinh(\di{\chi\over 2})V^{'\downarrow}& \nonumber \\
&{\rm where}~~~U^{'\downarrow}= \sqrt{m} \left[\exp(i\di{\varphi\over 2})
\cos(\di{\theta\over 2})U^{\downarrow}_0 - \exp(-i\di{\varphi\over 2})
\sin(\di{\theta\over 2})U^{\uparrow}_0 \right]& \nonumber
\end{eqnarray}

\vv \nin If the considered particle is a lepton, its spinors will be defined as above.
If that particle is a real photon, its 4-momentum $k$ is no more $\propto t$
(the latter 4-vector being time-like), but
should be considered as $k= E(t+z)$ (which is light-like), and its circular
polarizations will be defined as
\beq
\epsilon^{(\pm)} = \mp \di{1\over{\sqrt 2}}\left( x \pm i y \right)
\enq

\vv \nin As regards the latter formula, let us notice the useful identity

\beq
x =  \cos\varphi~X + \sin \varphi~Y -\tan(\di{\theta\over 2})\left(Z+Z'\right)
\enq

\vv \nin that allows us to rewrite the circular polarizations in the form

\beq
\epsilon^{(\pm)} = E^{(\pm)}\exp(\mp i \varphi) \pm\di{1\over
\sqrt{2}} \tan(\di{\theta\over 2})\left(\exp(-\chi)(t+z) -(T-Z)\right)
\enq

\vv \nin using the relation $T+ Z' = \exp(-\chi)(t+z)$.

\vv \nin If we make use of the gauge invariance of QED amplitudes, we see that we can
drop the term $\propto t+z \propto k$ in the above expression, and redefine
the circular polarizations of a real photon by

\beq
\epsilon^{(\pm)} =\exp(\mp i \varphi)\left(E^{(\pm)} + \xi^\star  (T-Z)
\right),~{\rm where}~~ 
\xi^\star = \mp \di{1\over
\sqrt{2}} \tan(\di{\theta\over 2})\exp(\pm i \varphi)
\label{polar} \enq

\vv \nin It is easy to check that the latter polarization 4-vectors are still
orthogonal to $k$. They are also orthogonal to $T - Z$.

\vv 

\section{On helicity properties of impact factors\protect\footnote{The definition of an impact factor is given in refs [2], [3].}}

\vv \nin To be specific, let us consider the impact factor corresponding to the QED 
subprocess lepton + $\gamma^{\star} \rightarrow $ lepton + jet. This 
impact factor is generally written as a sum of terms like 
$\bar{U}_3 {\cal T} U_1$, where $U_1$ is the bi-spinor of the incoming lepton, 
$U_3$ is the bi-spinor of a lepton of the same species pertaining to the final 
jet ; ${\cal T}$ is a 4X4 transition matrix which is a succession of products 
of lepton propagator and $\gamma$-matrices describing vertices (the other 
terms completing the impact factor correspond to possible exchanges between 
final particles). We will denote by 
$\lambda$ the helicity of the incoming lepton, by $\lambda$' that of the 
outgoing lepton and by $\Lambda$ the total helicity of the remaining particles 
in the final jet.

\vv \nin Let us perform a rotation of angle $\Phi$ around the Z-axis of the lab frame. 
This amounts to a redefinition of X and Y transverse axes. Under such a 
transformation, wave functions undergo the change 

\beq
\Psi_{\lambda_\psi} \rightarrow   \exp(-i \lambda_\psi \Phi) 
~\Psi'_{\lambda_\psi}
\enq

\noindent where $\Psi'_{\lambda_\psi}$ is the transformed wave function, and 
we must have 

\beq
J  =  \exp(i(\Lambda + \lambda' - \lambda)\Phi) ~ J'
\enq

\vv \nin  where $J'$ is obtained from $J$ by a simple redefinition of X and 
Y axes. 

\vv \nin Since denominators of propagators entering into $J$ possibly depend on 
azimutal angles only through relative combinations of those angles, they 
remain unchanged. These denominators will be left apart from the discussion regarding 
successive orders of approximation in the jet-like configuration. So, in the 
following, they should be considered as if they were simple constant 
coefficients.

\vv \nin Then, we may make the (very mild) assumption that $J$ is an analytical 
function of transverse components $p_{q x}$ and $p_{q y}$ of 4-momenta, and 
also an analytical function of the lepton mass. Instead of the above 
``linear'' transverse components, we may rather consider the ``circular'' 
combinations 

\beq
z_q = p_{q x} + i  p_{q y}~~~{\rm and}~~~z^{\star}_q = p_{q x} - i  p_{q y}
\label{leszpt} \enq

\vv \nin Such an assumption could be verified at least in the framework of jet-like 
kinematics where it could be possible to make the expansion 

\begin{eqnarray}
J\left(\left\{z_q\right\},\left\{z^{\star}_q\right\},m\right)
\approx J|_0 + m~ \di{{\pa J}\over{\pa m}}|_0 + m^2~\di{{\pa^2 J}\over{
(\pa m)^2}}|_0 + ... + \sum_q z_q ~\di{{\pa J}\over{\pa z_q}}|_0 
~~~~\nonumber \\  
+ \sum_q z^{\star}_q~ \di{{\pa J}\over{\pa z^{\star}_q}}|_0 
+ \di{1\over 2} \sum_q z^2_q ~\di{{\pa^2 J}\over{\pa z_q^2}}|_0 
+ \di{1\over 2} \sum_q z^{\star 2}_q ~\di{{\pa^2 J}
\over{\pa {z^{\star}_q}^2}}|_0 + \sum_{r<s} z_r z_s~
\di{{\pa^2 J}\over{\pa z_r \pa z_s}}|_0  \label{expansion} \\
+ \sum_{r<s} z^{\star}_r z^{\star}_s~
\di{{\pa^2 J}\over{\pa z^{\star}_r \pa z^{\star}_s}}|_0  
+ \sum_{r<s} z_r z^{\star}_s~\di{{\pa^2 J}\over{\pa z_r \pa z^{\star}_s}}|_0 
+ \sum_{r<s} z^{\star}_r z_s~\di{{\pa^2 J}
\over{\pa z^{\star}_r \pa z_s}}|_0 ~~~~~~~ \nonumber \\  
+ m~ \sum_q z_q ~\di{{\pa^2 J}\over{\pa m \pa z_q}}|_0 
+ m~ \sum_q z^{\star}_q ~\di{{\pa^2 J}\over{\pa m \pa z^{\star}_q}}|_0 
+ {\rm higher~order~terms}~~~~~ \nonumber 
\end{eqnarray}

\vv \nin Here, the symbol $|_0$ means that the corresponding quantity is taken at 
$m = 0$ and all $z_q$ and $z^{\star}_q$ equal to zero (except in denominators, 
as said above).  Of course, the leading order from which the expansion (\ref{expansion}) 
should start essentially depends on the subprocess considered. 

\vv \nin We may perform the same kind of expansion in both sides of eq (2). Taking 
into account the fact that 
\beq
z'_q = \exp(i\Phi)~z_q~~~{\rm and}~~~z^{'\star}_q = \exp(-i\Phi)~z^{\star}_q
\enq

\vv \nin we easily find, by simple identification of various terms, 
that we should have the constraints

\beq
\Lambda + \lambda' - \lambda = n - p \label{GRH}
\enq

\vv \nin on coefficients of the expansion corresponding to $p$ (for positive) 
factors $z_r$ and $n$ (for negative) factors $z^{\star}_r$. Such constraints 
automatically imply strong correlations between the relative values of 
helicities and the order of approximation. Let us consider some consequences.

\vv \nin {\bf 1)} For amplitudes without lepton-helicity-flip, we have 
$\lambda'=\lambda$, and thus $\Lambda = n - p$. For terms coming from the 
lepton mass only, we have $n=p=0$ and, therefore, $\Lambda = 0$. As seen in appendix \ref{mass-parity}, such amplitudes are even functions of $m$, so that terms with odd 
exponents of $m$ should be absent from the corresponding expansion.

\vv \nin We may then conclude the following.
\biz
\item For the elastic vertex $\ell + \gamma^{\star} \rightarrow \ell'$ 
where the final jet is made up of the final 
lepton only, the leading contribution is $J|_0$, and mass correction arise 
only at second order. In addition, since $n=p$, corrections from $z$'s and 
$z^{\star}$'s arise also at second order.

\item Regarding the process $\ell + \gamma^{\star} \rightarrow \ell' + 
\gamma$. 

\nin Now, we have $\Lambda_{\gamma} = \pm 1 = n-p$, or $n = p \pm 1$. This means 
that the leading order term cannot be $J|_0$ which is zero. The leading terms 
correspond to the cases $n=1,~p=0,~\Lambda=+1$ (terms $\propto p_x - i p_y$) 
and $n=0,~p=1,~\Lambda=-1$ (terms $\propto  p_x + i p_y$). Notice that 
corrections {\it from mass terms only} should cancel. The first mass 
corrections are of third order through terms involving $\pa^3 J/((\pa m)^2 
\pa z)$ ($n=1,~p=0,~\Lambda = 1$) or $\pa^3 J/((\pa m)^2 
\pa z^{\star})$ ($n=0,~p=1,~\Lambda = -1$). 

\item Regarding the process $\ell + \gamma^{\star} \rightarrow \ell' + 
\gamma_1 + \gamma_2$. 

We here have $\lambda_1 + \lambda_2 = n-p$. \\
For $\lambda_1=\lambda_2=+1$, 
then $n= p+2$ and the leading order is given by terms $\propto 
z^{\star}_r~z^{\star}_s$ 
($n=2,~p=0$). In that case, mass corrections cannot arise without $p_T$ 
corrections : the first corrections are of 4th order ($\sim m^2 ~z^{\star}_r~
z^{\star}_s$).\\
For $\lambda_1=\lambda_2=-1$, we have $p=n+2$ and leading contributions are 
$\propto z_r~z_s$. \\
For $\lambda_1= - \lambda_2$, we have $n=p$. In that case can we have a 
leading term $J|_0$ ? The answer seems to be : no. This may be related to the 
fact that 
\beq 
J \propto q_T~~~~{\rm as}~~~~q_T \rightarrow 0
\enq 

\nin as can be derived from the current conservation relation $q^\mu J_\mu = 0$. We thus conclude that the leading terms are either 
terms $\propto m^2$ or terms $\propto z_r~z^{\star}_s$.  
\eiz

\vv \nin {\bf 2)} Let us now turn to the case of lepton-helicity-flip 
amplitudes and take, for definiteness, $\lambda=-\lambda'= 1/2$. 
Then $\Lambda = n-p +1$. In addition, we know that such amplitudes are odd 
functions of $m$ and their corresponding expansions should not contain terms 
with even exponents of $m$.
\biz
\item For the elastic vertex, we get $n=p-1$ and the leading term is of  
second order $\propto m~z$ ($n=0,~p=1$).  

\item For $\ell + \gamma^{\star} \rightarrow \ell' + \gamma$, we have 
$\Lambda_{\gamma} = \pm 1 = n-p+1$ or $p = n+1 \mp 1$ and a leading term 
$\propto m$ is obtained for $\Lambda_{\gamma}=1$ when $n=p=0$, with next 
corrections of 3rd order ($\sim m^3,~\sim m~z_r~z^{\star}_s$). For 
$\Lambda_{\gamma}=-1$, $p = n + 2$ and the leading terms in this case are 
of 3rd order ($\sim m~z_r~z_s$). We see here 
the strong correlation between the helicity of the outgoing photon and that 
of the incoming lepton. 

\item  For the process $\ell + \gamma^{\star} \rightarrow \ell' + 
\gamma_1 + \gamma_2$, we have $\lambda_1+\lambda_2 = n-p +1$. \\
If $\lambda_1=\lambda_2=+1$, $n=p+1$ and the leading terms correspond to 
$n=1,~p=0$ and are $\propto m~z^{\star}$ (2nd order).\\
If $\lambda_1=-\lambda_2=+1$, $p=n+1$, and the leading terms correspond to 
$n=0,~p=1$ and are $\propto m~z$ (2nd order).\\ 
Finally, if $\lambda_1=\lambda_2=-1$, $p= n+3$, and this time, the leading 
terms are of order 4 and are $\propto m~z_q~z_r~z_s$ 
($n=0,~p=3$). In this case we can say that to 4th order the corresponding 
amplitude is zero. 

\vv \nin Here again, we see a strong correlation between the helicity of an outgoing 
photon and that of the incoming lepton, for lepton-helicity-flip amplitudes.
\eiz

\vv \nin So, it appears that most of helicity properties observed in specific calculations ({\bf [2-3]}) can be simply explained 
in such a formalism. For example, the maximum change in helicity implies a 
maximum value of $|n-p|$, and consequently leading terms of the expansion 
should be of higher order.

\vv \nin Anyway, for a former discussion about jet-like kinematics, we suggest to 
speak immediately about ``leading order calculations'', instead of saying 
that in such calculations only terms of first order in $m/E$ or $\theta's$ 
are kept while neglecting higher orders, which is not the case for the 
process $\ell + \gamma^{\star} \rightarrow \ell' + \gamma_1 + \gamma_2$
where amplitudes are of 2nd order. This would be more cautious and maybe more
clear.

\vv

\section{On analyticity of impact factors as regards angular variables} 

\vv \nin Let us notice that in jet-like kinematics, numerators of transition amplitudes are expanded in terms of small quantities mass $m$ and polar angles $\theta$. Thus, instead of using the variables (\ref{leszpt}) we will use the new ones 

\beq z_k = \theta_k \exp{(i \varphi_k)}~,~~{\rm and}~~~ z^\star_k \label{leszteta} \enq 

\vv \nin to express impact factors. If we extract from the transition amplitudes the phase factors $\exp(i \lambda_{\rm out} \varphi_{\rm out})$ for outgoing particles and $\exp(-i \lambda_{\rm in} \varphi_{\rm in})$ for ingoing particles, we assert that the remaining factor is, as regards angular variables, a function of the $z's$ and of the $z^\star$'s only. This can be understood in the following way. 

\vv \nin Spinors of the fundamental representation of spin 1/2 are written in the form

$$ u^\uparrow = \di{ \exp(- i \varphi/2) \over \sqrt{2}} 
\left( \begin{array} {c}
\cos (\theta/2) \\
\exp(i \varphi) \sin(\theta/2)  \end{array} \right)$$

\beq u^\downarrow = \di{ \exp( i \varphi/2) \over \sqrt{2}} 
\left( \begin{array} {c}
- \exp(-i \varphi) \sin(\theta/2)  \\
\cos(\theta/2)  \end{array} \right)\enq

\vv \nin and spinors of higher spin may be constructed from tensorial products of the above spinors. They are thus function of $\cos(\theta/2)$ and $\sin(\theta/2)$. More 
precisely, the latter functions are factorized with azimutal phase factors 
$\exp(i\varphi/2)$ and $\exp(-i\varphi/2)$. For spin up spinor we may 
extract the phase factor $\exp(-i\varphi/2)$ and the remaining is a linear 
(spinor) combination of $\cos(\theta/2)$ and $\sin(\theta/2)~\exp(i\varphi)$.
But the expansions 
\beq
\cos(\theta/2) = 1 - \theta^2/8 + ... = 1 - z z^{\star}/8 +...
\enq

\nin and 

\beq
\sin(\theta/2)~\exp(i\varphi) = ~\exp(i\varphi)\left( \theta/2 - 
\theta^3/48 + ...\right) = z/2 - 
z^2 z^{\star}/48 +...
\enq
 
\vv \nin clearly show that they are analytic functions of the variables $z$ and $z^{\star}$ as defined by (\ref{leszteta}). An analogous 
result holds for the spin down spinor. As for photons, we have (applying 
gauge invariance) 

\beq
\epsilon^{(\lambda)} = \exp(-i\varphi)\left(E^{(\lambda)} - \di{\lambda
\over{\sqrt{2}}} \exp(i\lambda \varphi)~\tan(\theta/2)~(T-Z)\right)
\enq

\nin Then, for example, 
\beq
\tan(\theta/2)~\exp(i\varphi) =\exp(i\varphi)\left( \theta/2 +
\theta^3/24 + ...\right) = z/2 + 
z^2 z^{\star}/24 +...
\enq

\nin which is also analytic in $z$ and $z^{\star}$.

\vv \nin Scalar products between 4-momenta and 
polarization 4-vectors can also be considered as analytic functions of $z$ 
and $z^{\star}$. As for the scalar product of two 4-momenta $p$ and $p'$ 
it involves the scalar product of unitary 3-vectors 
\begin{center}
$\vec{n} = 
(\sin(\theta)~\cos(\varphi),~\sin(\theta)~\sin(\varphi),~\cos (\theta))$ \\
\end{center}
and  
\begin{center}
$\vec{n'} = 
(\sin(\theta')~\cos(\varphi'),
~\sin(\theta')~\sin(\varphi'),~\cos (\theta'))$   
\end{center}
i.e :
\beq 
\vec{n}.\vec{n'} = \cos(\theta)~ \cos(\theta') + \sin(\theta)~\sin(\theta')
\cos(\varphi -\varphi')
\enq

\vv \nin Transforming $\cos(\varphi -\varphi')$ into the sum of exponentials 
$\exp(i(\varphi -\varphi'))$ and $\exp(-i(\varphi -\varphi'))$ we see 
that such scalar products involve combinations of 
\begin{center}
$\cos \theta$, $\cos \theta'$, 
$ \sin \theta~\exp(\pm i \varphi)$ and $ \sin \theta'~\exp(\pm i \varphi')$
\end{center}

\nin We can write 
\begin{eqnarray}
\cos \theta = 1 -z z^{\star}/2 + ... \nonumber  \\
\cos \theta' = 1 -z' z^{'\star}/2 + ... \nonumber  \\
\sin \theta~\exp( i \varphi) = z - z^2 z^{\star}/6 + ... \\
\sin \theta~\exp(- i \varphi) = z^{\star} - z z^{\star 2}/6 + ... \nonumber \\
\sin \theta'~\exp( i \varphi') = z' - z^{'2} z^{'\star}/6 + ... \nonumber \\
\sin \theta'~\exp(- i \varphi') = z^{'\star} - z' z^{'\star 2}/6 + ... 
\nonumber 
\end{eqnarray}

\vv \nin which demonstrates their analyticity. A scalar product of a 4-momentum $p_n$ and a polarization 4-vector of a final 
photon is written as 
\begin{eqnarray}
\epsilon^{(\lambda)\star}.p_n = \di{\lambda\over{\sqrt{2}}}\left( p_{nT} 
\exp{(i\lambda[\phi-\phi_n])}  - p_{n+} \tan(\theta/2) \right) \\ \nonumber
= \di{\lambda\over{\sqrt{2}}} \exp(i\lambda \phi)\left( |\vec{p_n}| 
\sin \theta_n 
\exp(-i\lambda\phi_n) -   p_{n+} \tan(\theta/2) \exp(-i\lambda \phi) 
\right) 
\end{eqnarray}

\vv \nin As seen before, $\sin \theta_n~\exp(-i\lambda\phi_n)$ and $\tan(\theta/2) 
\exp(-i\lambda \phi)$ are analytic functions respectively, of $z_n$ and 
$z^{\star}_n$, and of $z_\gamma$ and $z^{\star}_\gamma$. So, such scalar 
products are analytic functions of the $z$ and $z'$ involved and of their 
conjugates. This, in fact 
was to be expected since scalar products of ordinary 3-vectors are 
constructed from tensorial products of components of spinors pertaining to 
the fundamental representation of spin 1/2 of the rotational group. But we 
have seen that apart from global phase factors $\exp( i \lambda \varphi)$ 
that we factorize, they are analytic in the $z$ and $z^{\star}$.

\vv \nin Finally, when multiplying all these 
functions we get surely an analytic expression in $z$ and $z^{\star}$, and 
thus (the numerator of) the impact factor should be analytic in $z$ and $z^{\star}$. 

\vv \nin To 
illustrate this, let us consider some examples. 
\biz

\item Some helicity amplitudes $A(\lambda_{\gamma_1},~\lambda_{\gamma_2},~
\lambda_{\ell_1},~\lambda_{\ell_2})$ for the Compton effect $\ell_1 +\gamma_1 
\rightarrow \ell_2 +\gamma_2$. Apart from coupling constant, we have 
\begin{eqnarray*}
A(+,+,\downarrow,\uparrow) = -\exp(-i\varphi_{\gamma_1}+i\varphi_{\gamma_2}
+i\varphi_{\ell_1}/2-i\varphi_{\ell_2}/2)
\left(\di{{2\cos^3(\theta/2)}\over{1-\beta'\sin^2(\theta/2)}}\right) 
\end{eqnarray*}
\begin{eqnarray*}
A(+,+,\uparrow,\downarrow,) = \exp(-i\varphi_{\gamma_1}+i\varphi_{\gamma_2}
-i\varphi_{\ell_1}/2-i\varphi_{\ell_2}/2) . \\
\left(\di{{2m\exp(i\varphi_{\ell_2})\sin(\theta)\cos(\theta/2)}
\over{\sqrt{s}(1-\beta'\sin^2(\theta/2))}}\right) 
\end{eqnarray*}
\begin{eqnarray*}
A(+,-,\downarrow,\uparrow,) =- \exp(-i\varphi_{\gamma_2}-i\varphi_{\gamma_1}
+i\varphi_{\ell_1}/2+i\varphi_{\ell_2}/2) . \\
\left(\di{{2m\exp(i\varphi_{\ell_2})\sin^3(\theta/2)}
\over{\sqrt{s}(1-\beta'\sin^2(\theta/2))}}\right) 
\end{eqnarray*}

\vv \nin where $\theta$ is the emission angle of $\ell_2$ and $\beta'=1-m^2/s$. Of 
course, we usually take all $\varphi$'s =0.

\item Vertex pion-nucleon with a simple Yukawa coupling. Amplitudes 
$A(\lambda_3,\lambda_1)$ are $\propto \bar{U}_3 \gamma_5 \bar{U}_1$. Apart 
from factors depending on mass and energies and omitting denominators, we find
\begin{eqnarray*}
A(\uparrow,\uparrow) \sim \exp(i\varphi_3/2-i\varphi_1/2)
\left(\cos(\theta_1/2)\cos(\theta_3/2)+\right. \\
\left. \sin(\theta_1/2)\exp(-i\varphi_1)\sin(\theta_1/2)
\exp(-i\varphi_3) \right)
\end{eqnarray*}
\begin{eqnarray*}
A(\uparrow,\downarrow) \sim 
\exp(i\varphi_3/2+i\varphi_1/2)\left(\cos(\theta_1/2)
\sin(\theta_3/2)\exp(-i\varphi_3)- \right.\\
\left. \cos(\theta_3/2)\sin(\theta_1/2)
\exp(-i\varphi_1)\right) 
\end{eqnarray*}

\item Pion-Nucleon scattering with Yukawa coupling $\Pi(q) +N(p_1)\rightarrow 
\Pi(k) + N(p_3)$. A very simple calculation shows that the numerator of the 
amplitude is proportional to $\bar{U}_3 \hat{k} \bar{U}_1$.  
\begin{eqnarray*}
A(\uparrow,\uparrow) \sim 
\exp(i\varphi_3/2-i\varphi_1/2)\left( \exp(\alpha_1/2+\alpha_3/2) \left(
\cos(\theta_3/2) k_{-} \right.\right.  \\
\left. \left. -\sin(\theta_3/2)\exp(-i\varphi_3)(k_x +i k_y)
\right. \right.\\
\left. \left.+ \exp(-\alpha_1/2-\alpha_3/2) \left( \cos(\theta_3/2) k_{+} +
\sin(\theta_3/2)\exp(-i\varphi_3)(k_x+ik_y) \right)\right.\right)  
\end{eqnarray*}
with $k_{+} = k_0 + k_z$, $k_{-} = (m^2_{\pi}+ k^2_{T})/k_{+}$, $\theta_1=0$ 
(with $\theta_1 \neq 0$, we get a longer result but the general structure remains the same). 
\eiz

\vv \nin From these examples, we see that once the phase factor $\exp\left(i \sum 
(\lambda_{out}\varphi_{out} - 
\lambda_{in}\varphi_{in})\right)$ has been extracted, the remaining factor 
is indeed a functional of the $z$'s and $z^{\star}$'s of all the particles 
entering into the reaction, through simple trigonometric functions that admit 
a Taylor expansion relatively to $z$ and $z^{\star}$'. We may consider this as a very general result, independent on the process 
and on energies as well, that could be ascribed to the way we calculate 
transition amplitudes in the framework of Quantum Field theory.

\vv \nin So, as regards angular variables, impact factors take on the general 
form
\beq
J \sim \exp\left(i \sum (\lambda_{out}\varphi_{out} - 
\lambda_{in}\varphi_{in})\right)~F(\left\{z\right\},\left\{z^{\star}\right\})
\enq

\vv \nin As is expected from the common belief in analyticity of 
S-matrix elements, the function $F$ can be expanded in powers, with positive 
integer exponents, of $z$ and 
$z^{\star}$, at least in the domain $|z| \ll 1$. Indeed, this appears to be  
the case for all the $F$ we are dealing with. This is a sufficient condition for proving the helicity properties of impact factors found in section 2.

\vv \vv

\section{Exact results about the strict forward direction where all $|z|$'s = 0}

\vv \nin As a general remark, we remind that denominators of propagators are left 
apart from the discussion.

\subsection{The subprocess 
$\ell_1(p_1) + \gamma^{\star}(q) \rightarrow \ell_3(p_3) + (N-1) 
{\rm{\bf real~photons}}(k_j)$} 

\vv
\subsubsection{Impact factors}

\vv \nin Impact factors are projection onto the light-like 4-vector $T-Z$. A generic 
form for numerators of these impact factors is 
\beq 
{\cal N} = \bar{U}_3 A \g(T-Z) B  U_1
\label{FW1} \enq

\nin where the matrix $B$ ($B$ for before) is given by 

\begin{eqnarray}
& B= [m+\g(p_1-k_1-k_2-\cdots-k_{p-1})] \g(\ep^{\star}_{p-1}) 
[m+\g(p_1-k_1-k_2-\cdots & \nonumber \\
& -k_{p-2})]\g(\ep^{\star}_{p-2})\cdots[m+\g(p_1-k_1)]\g(\ep^{\star}_1) &
\label{FW2} \end{eqnarray}

\nin and the matrix $A$ ($A$ for after) given by
\begin{eqnarray}
& A=  \g(\ep^{\star}_N)[m+\g(p_3+k_N)]\g(\ep^{\star}_{N-1})
[m+\g(p_3+k_N+k_{N-1})] \g(\ep^{\star}_{N-2})\cdots& \nonumber \\
& \times \g(\ep^{\star}_{p+1})
[m+\g(p_3+k_N+k_{N-1}+\cdots+k_{p+1})] &
\label{FW3} \end{eqnarray}

\nin where the notations $\g(v) = \g_{\mu}v^{\mu}$ and $k_p = -q$ ($q$ is the 
4-momentum of the virtual photon) have been used.
In the above amplitude, the vertex with the virtual photon is inserted 
at the $p$-th rank, after the emission of $n^b_{\g} = p-1$ real photons 
($b$ for before) and 
before the emission of $n^a_{\g}= N-p$ ($a$ for after) real photons, from 
the lepton line. Let us also denote by $n_{\g} =n^b_{\g}+n^a_{\g}=N-1$ 
the total number of real photons emitted.

\vv \nin For the forward direction, we have 
\begin{eqnarray}
& p_3 \equiv {\cp}_3 = E_3 T + |\vec{p}_3| Z & \nonumber \\
&p_1 \equiv {\cp}_1 = E_1 T + |\vec{p}_1| Z~~ (\rm{this~last~one~is~ exact~ 
anyway}) & \label{FW4} \\
& k_j \equiv {\ck}_j = \omega_j( T + Z) & \nonumber \\  
&\epsilon^{\star}_j \equiv E^{\star}_j \equiv \exp(i\lambda_j \varphi_j) 
E^{(\lambda_j) \star} ~~~{\rm{with}}~~~ E^{(\pm)} = \mp 
\di{1\over{\sqrt{2}}}
( X \pm i Y) & \nonumber
\end{eqnarray}

\nin and the generic form ${\cal N}$ becomes 
\beq 
{\cal N} = \bar{\uz}_3 \az \g(T-Z) 
\bz \uz_1
\label{FW5} \enq

\nin where the symbol ``$\circ$" above letters means that we are taking all quantities 
for all $\theta$'s $=0$. Using Dirac equation and the above forms (\ref{FW4}) of 4-vectors we get 

\begin{eqnarray}
& \bar{ \uz}_3 \g(E^{\star}_N)[m+\g(\cp_3+\ck_N)]
\g(E^{\star}_{N-1})[m+\g(\cp_3+\ck_N+\ck_{N-1})] & \nonumber \\
& = \bar{ \uz}_3 \g(E^{\star}_N)
\g(\ck_N)\g(E^{\star}_{N-1})
[m+\g(\cp_3+\ck_N+\ck_{N-1})] & \label{FW6} \\
& = \bar{ \uz}_3 \g(E^{\star}_N)
\g(\ck_N)\g(E^{\star}_{N-1})
[m+\g(\cp_3)] =- 2 \cp_3\cdot \ck_N  \bar{ \uz}_3 
\g(E^{\star}_N)\g(E^{\star}_{N-1})& \nonumber
\end{eqnarray}

\nin Next, 

\begin{eqnarray}
& \bar{ \uz}_3 \g(E^{\star}_N) 
\g(E^{\star}_{N-1})\g(E^{\star}_{N-2})
[m+\g(\cp_3+\ck_N+\ck_{N-1}+\ck_{N-2})] \g(E^{\star}_{N-3}) & \nonumber \\ 
& \times [m+\g(\cp_3+\ck_N+\ck_{N-1}+\ck_{N-2}+\ck_{N-3})] \equiv 
\bar{ \uz}_3 \g(E^{\star}_N) 
\g(E^{\star}_{N-1})\g(E^{\star}_{N-2}) & \nonumber \\
& \times\g(\ck_N+\ck_{N-1}+\ck_{N-2}) \g(E^{\star}_{N-3}) [m+\g(\cp_3)]  
 & \label{FW7} \\
&= - 2 \cp_3 \cdot (\ck_N+\ck_{N-1}+\ck_{N-2})  \, \bar{ \uz}_3 \g(E^{\star}_N)\g(E^{\star}_{N-1})
\g(E^{\star}_{N-2}) \g(E^{\star}_{N-3}) & \nonumber 
\end{eqnarray}

\nin and so on. Then, either $n^a_{\g}= N-p$ is even and we can go on the 
reduction until ($N
-p-1$ is odd)
\begin{eqnarray}
& \bar{ \uz}_3 \az \equiv (- 2 \cp_3\cdot \ck_N)(- 2 
\cp_3\cdot (\ck_N+\ck_{N-1}+\ck_{N-2}))\cdots 
(- 2 \cp_3\cdot (\ck_N+\ck_{N-1}+& \nonumber \\
& \ck_{N-2}+
\cdots+\ck_{p+2}))  
~\bar{ \uz}_3 \g(E^{\star}_N)\g(E^{\star}_{N-1})
\g(E^{\star}_{N-2})\cdots \g(E^{\star}_{p+1})& \label{FW8}
\end{eqnarray}

\nin or $n^a_{\g}$ is odd ($N-p-1$ is even) and we must stop the reduction at  

\begin{eqnarray}
& \bar{ \uz}_3 \az \equiv (- 2 \cp_3\cdot \ck_N)
(- 2 \cp_3\cdot (\ck_N+\ck_{N-1}+\ck_{N-2}))\cdots 
(- 2 \cp_3\cdot (\ck_N+\ck_{N-1}+& \nonumber \\
& +\ck_{N-2}+\cdots+\ck_{p+3}))  
~\bar{ \uz}_3 \g(E^{\star}_N)\g(E^{\star}_{N-1})
\g(E^{\star}_{N-2})\cdots \g(E^{\star}_{p+1}) & \label{FW9}  \\
& \times \g(\ck_N+\ck_{N-1}+\cdots+
\ck_{p+1}) & \nonumber
\end{eqnarray}

\vv \nin On the side of the incoming lepton, we can perform the same kind of reduction. 
We thus get 
\begin{eqnarray}
& \bz \uz_1 \equiv (2 \cp_1\cdot \ck_1)(2 \cp_1\cdot 
(\ck_1+\ck_2+\ck_3))\cdots 
(2 \cp_1\cdot (\ck_1+\ck_2+\cdots+\ck_{p-2})) & \nonumber \\ 
& \times\g(E^{\star}_{p-1})\g(E^{\star}_{p-2})\cdots \g(E^{\star}_1) 
\uz_1 & \label{FW10} 
\end{eqnarray}

\nin if $n^b_{\g}= p-1$ is even and   
\begin{eqnarray}
& \bz \uz_1 \equiv - (2 \cp_1\cdot\ck_1)(2 \cp_1\cdot 
(\ck_1+\ck_2+\ck_3))\cdots 
(2 \cp_1\cdot (\ck_1+\ck_2+\cdots+\ck_{p-3})) &\nonumber \\ 
&\times\g(\ck_1+\ck_2+\cdots+\ck_{p-1})
\g(E^{\star}_{p-1})\g(E^{\star}_{p-2})\cdots \g(E^{\star}_1) 
\uz_1 & \label{FW11}
\end{eqnarray}

\nin if $n^b_{\g}= p-1$ is odd.

\vv \nin Let us now consider separately the cases where  $n_{\g}$ is even or odd.

\vv \nin {\bf 1) \und{$n_{\g}$ is odd.}}\\

\nin We have then the only two possibilities : either $n^a_{\g}$ is even and  
$n^b_{\g}$ is odd, or $n^a_{\g}$ is odd and $n^b_{\g}$ is even. Accordingly, 
we are led to expressions like 
\begin{eqnarray}
&(- 2 \cp_3\cdot \ck_N)(- 2 \cp_3\cdot (\ck_N+\ck_{N-1}+\ck_{N-2}))\cdots 
(- 2 \cp_3\cdot (\ck_N+\ck_{N-1}+& \nonumber \\
& +\ck_{N-2}+\cdots+\ck_{p+2}))& \nonumber  \\
& \times \bar{ \uz}_3 \g(E^{\star}_N)\g(E^{\star}_{N-1})
\g(E^{\star}_{N-2})\cdots \g(E^{\star}_{p+1})\g(T-Z) &  \label{odd4-1} \label{FW12} \\
& \g(\ck_1+\ck_2+\cdots+\ck_{p-1}) \g(E^{\star}_{p-1})\g(E^{\star}_{p-2})
\cdots \g(E^{\star}_1) \uz_1 & \nonumber  \\
& \times (-) (2 \cp_1\cdot \ck_1)(2 \cp_1\cdot (\ck_1+\ck_2+\ck_3))\cdots 
(2 \cp_1\cdot (\ck_1+\ck_2+\cdots+\ck_{p-3})) & \nonumber
\end{eqnarray}

\nin for $n^a_{\g}$ even and  $n^b_{\g}$ odd, and 

\begin{eqnarray}
& (- 2 \cp_3\cdot \ck_N)
(- 2 \cp_3\cdot (\ck_N+\ck_{N-1}+\ck_{N-2}))\cdots 
(- 2 \cp_3\cdot (\ck_N+\ck_{N-1}+& \nonumber \\
&+\ck_{N-2}+\cdots+\ck_{p+3})) & \nonumber \\
& \bar{ \uz}_3 \g(E^{\star}_N)\g(E^{\star}_{N-1})
\g(E^{\star}_{N-2})\cdots \g(E^{\star}_{p+1}) \g(\ck_N+\ck_{N-1}+
\cdots+\ck_{p+1}) &  \label{FW13} \\
& \times \g(T-Z)\g(E^{\star}_{p-1})\g(E^{\star}_{p-2})\cdots \g(E^{\star}_1) 
\uz_1 & \nonumber \\
& \times (2 \cp_1\cdot \ck_1)(2 \cp_1\cdot (\ck_1+
\ck_2+\ck_3))\cdots 
(2 \cp_1\cdot (\ck_1+\ck_2+\cdots+\ck_{p-2})) & \nonumber
\end{eqnarray}

\nin for $n^a_{\g}$ odd and  $n^b_{\g}$ even. But 
\beq
\g(\ck_1+\ck_2+\cdots+\ck_{p-1}) =(\omega_1+\cdots+\omega_{p-1}) \g(T+Z)
\label{FW14} \enq
\nin and 
\beq
\g(\ck_N+\ck_{N-1}+\cdots+\ck_{p+1})=(\omega_N+\cdots+\omega_{p+1}) \g(T+Z)
\label{FW15} \enq

\vv \nin Thus the generic form should be proportional to 
\begin{eqnarray}
& \bar{ \uz}_3 \g(E^{\star}_N)
\cdots \g(E^{\star}_{p+1})\g(E^{\star}_{p-1})
\cdots \g(E^{\star}_1) \g(T-Z) \g(T+Z) \uz_1 &
\label{FW16} \end{eqnarray}

\nin in the first case or 
\begin{eqnarray}
& \bar{ \uz}_3 \g(E^{\star}_N)
\cdots \g(E^{\star}_{p+1})\g(E^{\star}_{p-1})
\cdots \g(E^{\star}_1) \g(T+Z) \g(T-Z)\uz_1 &
\label{FW17} \end{eqnarray}

\nin in the second case, or, equivalently, proportional to  
\begin{eqnarray}
\bar{ \uz}_3 \g(E^{\star}_N)
\cdots \g(E^{\star}_{p+1})\g(E^{\star}_{p-1})
\cdots \g(E^{\star}_1)\left(1\mp 2\lambda_1 \g_5\right)\uz_1 
\label{FW18} \end{eqnarray}

\nin Such an expression is zero if 
\biz
\item two successive $E^{\star}_j$ are the same

\item $\lambda_{\g_1} = -2 \lambda_1$
\eiz

\nin Thus, in order to obtain a non-zero result, we must have
\beq
\lambda_{\g_1} = -\lambda_{\g_2}=\lambda_{\g_3}= \cdots = \lambda_{\g_{N-2}}
=- \lambda_{\g_{N-1}} = 2 \lambda_1
\label{FW19} \enq  

\nin and the studied term is then proportional to 
\beq
\bar{ \uz}_3 \g(E^{\star}_N)
\left(1\mp 2\lambda_1 \g_5\right)\uz_1 
\label{FW20} \enq

\nin which is non-zero only for $\lambda_{\g_N}= -2 \lambda_3 = 2 \lambda_1$. 
As a conclusion, \und{when $n_\g$ is odd}, \und{only HNC amplitudes have 
non-zero values}\footnote{HNC means 
: helicity non-conserving, and HC : helicity conserving}. This was 
to be expected from the general rule (\ref{GRH}) :  
$\Lambda + \lambda_3 - \lambda_1 = n-p$ 
that here gives ($n=p=0$) 
$\Lambda = \lambda_1 - \lambda_3$. But when $n_\g$ is odd, $\Lambda$ is 
necessarily odd and so should also be the difference $\lambda_1 - \lambda_3$. 
Obviously, this can be achieved only for HNC amplitudes.

\vv \nin Moreover, it can be easily shown that (\ref{FW20}) is proportional to the lepton 
mass $m$\footnote{Spinors are normalized according to $\bar{U} U = 2 m$}. As 
for scalar 
products of 4-vectors appearing in (\ref{FW12})  or (\ref{FW13}) we have, for example,  
\beq
\cp_3\cdot \ck_j = \omega_j \cp_{3-} = m^2 \omega_j /\cp_{3+}~,~~~
\cp_1\cdot \ck_r = m^2 \omega_r /\cp_{1+}  
\label{FW21} \enq

\nin so that the resulting amplitude (\ref{FW12}) is proportional to 
\beq
(m^2)^{(N-p)/2} (m^2)^{(p-2)/2} m = m^{n_\g}
\label{FW22} \enq

\vv \nin It is easy to show that the same result applies for the amplitude (\ref{FW13}). 
For instance, for $n_\g=1$, only the spin-flip amplitude is a priori non-zero 
for the forward direction and its numerator should be $\propto m$, which is 
indeed the case. We also notice that we here find again that HNC amplitudes 
should be odd functions of $m$.

\vv \nin {\bf 2) \und{$n_{\g}$ is even.}}

\vv \nin In this case, we have the two possibilities : $n^a_{\g}$ and $n^b_{\g}$ 
both even or $n^a_{\g}$ and $n^b_{\g}$ both odd. 
\bigskip

\nin a) $n^a_{\g}$ and $n^b_{\g}$ both even.

\vv \nin The generic form is then 
\begin{eqnarray}
&(- 2 \cp_3\cdot \ck_N)\cdots(- 2 \cp_3\cdot (\ck_N+\ck_{N-1}+
+\cdots+\ck_{p+2}))& \nonumber  \\
& \times \bar{ \uz}_3 \g(E^{\star}_N)
\cdots \g(E^{\star}_{p+1})\g(T-Z) 
\g(E^{\star}_{p-1})\cdots \g(E^{\star}_1) \uz_1 &   \label{FW23} \\
& \times (2 \cp_1\cdot \ck_1)\cdots 
(2 \cp_1\cdot (\ck_1+\ck_2+\cdots+\ck_{p-2})) & \nonumber
\end{eqnarray}

\nin The above matrix element is non-zero only if 
\beq
\lambda_{\g_1} = 2 \lambda_1= -\lambda_{\g_2}=\lambda_{\g_3}= \cdots = 
\lambda_{\g_{N-1}} = - \lambda_{\g_N} = 2 \lambda_3
\label{FW24} \enq  

\vv \nin and thus, all analogous amplitudes are non-zero if 
\beq
\sum_{j\neq p} \lambda_{\g_j} = 0 ~~~~~~{\rm and}~~~~~~~\lambda_1= \lambda_3
\label{FW25} \enq

\nin Under these conditions, the matrix element is proportional to the following 
one 
\beq
\bar{ \uz}_3 (\lambda_1) \g(T-Z)\uz_1 (\lambda_1)
\label{FW26} \enq

\nin which can be shown to be of order zero in $m$. Consequently, this case a) 
corresponds to terms in \und{HC amplitudes} that are proportional to 
\beq
(m^2)^{(N-p)/2} (m^2)^{(p-1)/2} m = m^{n_\g}
\label{FW27}\enq

\vv \nin b) $n^a_{\g}$ and $n^b_{\g}$ both odd.

\vv \nin Now, the generic form is 
\begin{eqnarray}
&(- 2 \cp_3\cdot \ck_N)\cdots(- 2 \cp_3\cdot (\ck_N+\ck_{N-1}+
+\cdots+\ck_{p+3}))& \nonumber  \\
& \times \bar{ \uz}_3 \g(E^{\star}_N)
\cdots \g(E^{\star}_{p+1}) \g(\ck_N+\ck_{N-1}+
\cdots+\ck_{p+1}) & \nonumber \\
&\g(T-Z) \g(\ck_1+\ck_2+\cdots+\ck_{p-1}) \g(E^{\star}_{p-1})\cdots \g(E^{\star}_1) \uz_1 &  \label{FW28} \\
& \times (-) (2 \cp_1\cdot \ck_1)\cdots 
(2 \cp_1\cdot (\ck_1+\ck_2+\cdots+\ck_{p-3})) & \nonumber
\end{eqnarray}

\nin But because of (\ref{FW13}), (\ref{FW14}), (\ref{FW15}) and since $\g(T+Z)\g(T-Z)\g(T+Z)= 2 \g(T+Z)$, 
this amplitude is proportional to 
\beq
\bar{ \uz}_3 \g(E^{\star}_N)
\cdots \g(E^{\star}_{p+1}) \g(E^{\star}_{p-1})\cdots \g(E^{\star}_1) 
\g(T+Z) \uz_1 
\label{FW29} \enq

\nin Here again, we find that this term is non-zero only if conditions (\ref{FW25}) are 
satisfied, and we are led to the simple matrix element 
\beq
\bar{ \uz}_3(\lambda_1) \g(T+Z) \uz_1(\lambda_1) 
\label{FW30} \enq

\nin which is proportional to $m^2$. Therefore, the terms corresponding to 
that case b) are associated with HC amplitudes and are proportional to 
\beq
(m^2)^{(N-p-1)/2} (m^2)^{(p-2)/2} m^2 = m^{n_\g}
\label{FW31} \enq

\nin So, as a general conclusion for the subprocess under study, we may state that 
impact factors in the forward direction are all proportional to $m^{n_\g}$. 
Moreover, in this limit, only HC amplitudes survive if $n_\g$ is even 
while only HNC amplitudes survive if $n_\g$ is odd. This is in complete 
agreement with the above-mentioned general rule 
$\Lambda + \lambda_3 - \lambda_1 
= n-p$ for $n=p=0$ :  
\biz
\item if $n_\g$ is even, $\Lambda$ is even and we must have 
$\lambda_3 = \lambda_1$, and then $\Lambda=0$ ;

\item if $n_\g$ is odd, $\Lambda$ is odd and we must have 
$\lambda_3 = - \lambda_1$, $\Lambda= 2 \lambda_1 $. 
\eiz

\subsubsection{Other projections}

\vv \nin {\bf 1) $J_{-}$ components}

\vv \nin This component is the projection of the current onto $T+Z$. It is clear from 
(\ref{FW14}), (\ref{FW15}) and (\ref{FW28}) that replacing $T-Z$ by $T+Z$ yields a zero result. Thus, 
as far as numerators are concerned, the component $J_{-}$ is strictly zero 
in the forward direction configuration if $n_\g$ is odd. If $n_\g$ is even, 
we a priori get a non-zero result if $n^a_\g$ and $n^b_\g$ are both even. 
This concerns only HC amplitudes and the corresponding amplitude is 
proportional to (\ref{FW30}) which itself is $\propto m^2$. Finally, $J_{-}$ is 
proportional to $m^{n_\g+2}$. A similar result may be found in 
another context.

\vv \nin The relation that expresses current conservation   
\beq
q^{\mu}J_{\mu} = 0 = J_{-} q_{+}/2 + J_{+} q_{-}/2
-\vec{q}_T \cdot \vec{J}_T  
\label{FW32} \enq

\nin for the current $J$ describing the subprocess, would give, for the forward 
direction (keeping a priori denominators unchanged...), 
\beq
 J_{-} = J\cdot (T+Z) = - q_{-} J_{+}/q_{+}
\label{FW33} \enq

\nin But in this limit $ q_{-} =  {\cp}_{3 -} - {\cp}_{1 -} = m^2 (1/{\cp}_{3 +}-
1/{\cp}_{1 +})$. Thus, we find that in this configuration 
\beq 
J_{-}/ J_{+} \propto m^2/s
\label{FW34} \enq

\nin for any set of helicities.

\vv \nin However, one should be aware of the fact that current conservation involves 
the whole set of factors in the current, not only numerators but 
denominators as well. Thus, taking $|\vec{q}_T| =0$ in (\ref{FW32}) certainly involves 
approximate forms of denominators. This explains why, for instance, in the 
former derivation (where denominators are left aside) we got a zero value 
for $J_{-}$ when $n_\g$ is odd while $J_{+}$ is not zero in \mbox{that case :} 
this means that in the kind of approximation we have in mind 
where denominators 
are kept more or less 
``exact'',  formula (\ref{FW32}) should be considered with great care.

\vv \nin {\bf 2) $J_T$ components}

\vv \nin Transverse components are projections onto $E^{(\pm)}$. Here too, we will 
discuss separately the cases $n_\g$ even and $n_\g$ odd.

\vv \nin {\bf a) \und{$n_\g$ even}}

\vv \nin Since $\g(T+Z)\g(E^{(\pm)})\g(T+Z) =0$, the case ``$n^a_\g$ odd $n^b_\g$ odd'' 
is excluded. For the case ``$n^a_\g$ even $n^b_\g$ even'', the generic matrix 
element is analogous to (\ref{FW23}) with $E^{(\pm) \star}$ replacing $T-Z$. Then, we should 
have 
\begin{eqnarray}
&\lambda_{\g_1} = 2 \lambda_1= -\lambda_{\g_2}=\lambda_{\g_3}= \cdots = 
\lambda_p ~(\rm{virtual~photon}) = & \nonumber \\
& \cdots = -\lambda_{\g_{N-1}} = \lambda_{\g_N} = - 2 \lambda_3 &
\label{FW35} \end{eqnarray}  

\nin This case is thus associated with HNC amplitudes and the remaining matrix 
element is 
\beq
\bar{ \uz}_3(-\lambda_1) \g(E^{\star}_N) 
\uz_1(\lambda_1) 
\label{FW36} \enq

\nin which can be easily found to be $\propto m$. In conclusion, if $n_\g$ is 
even, only HNC amplitudes survive and are $\propto m^{n_\g+1}$. Moreover, 
we have the correlation $\lambda_p = 2 \lambda_1= - 2 \lambda_3$.

\vv \nin {\bf b) \und{$n_\g$ odd}}

\vv \nin Let us consider the case where $n^a_\g$ is even and $n^b_\g$ odd. Then, 
if we adapt (\ref{FW12}) to this case, the generic matrix element is 
\begin{eqnarray}
&(- 2 \cp_3\cdot \ck_N)\cdots (- 2 \cp_3\cdot (\ck_N+\ck_{N-1}
+\cdots+\ck_{p+2}))& \nonumber  \\
& \times \bar{ \uz}_3 \g(E^{\star}_N)
\cdots \g(E^{\star}_{p+1})\g(E^{\star}_{p}) \g(E^{\star}_{p-1})
\cdots \g(E^{\star}_1) \g(T+Z) \uz_1 &  \label{FW37} \\
& \times (-) (2 \cp_1\cdot \ck_1)\cdots 
(2 \cp_1\cdot (\ck_1+\ck_2+\cdots+\ck_{p-3})) & \nonumber
\end{eqnarray}

\nin This matrix element is non-zero only if 
\begin{eqnarray}
&\lambda_{\g_1} = 2 \lambda_1= -\lambda_{\g_2}= \cdots = 
- \lambda_p = \cdots = \lambda_{\g_{N-1}} = - \lambda_{\g_N} =  2 \lambda_3 &
\label{FW38} \end{eqnarray}  

\nin It thus concerns only HC amplitudes and after reduction, the remaining 
matrix element is proportional to (\ref{FW30}). We can conclude that when 
$n_\g$ is odd, only HC amplitudes are non-zero for the forward direction 
and that they are proportional to  
\beq
(m^2)^{(N-p)/2} (m^2)^{(p-2)/2} m^2 = m^{n_\g+1}
\label{FW39}\enq

\vv \nin So, as a general conclusion for this subprocess, we may state that the 
transverse components of its current for the forward direction are 
$\propto m^{n_\g+1}$. Moreover, if $n_\g$ is even only HNC amplitudes 
survive (and $\lambda_p = 2 \lambda_1$) while if $n_\g$ is 
odd only HC amplitudes survive (and $\lambda_p = -2 \lambda_1$).

\vv \nin Notice that some of these results could be derived from the general constraint 
$\Lambda+\lambda_p = \lambda_1 -\lambda_3$ where the helicity $\pm$ 
of the virtual photon should be now included. 

\vv \nin Indeed, if $n_\g$ is even, $\Lambda+\lambda_p$ is odd, and we must have 
$\lambda_1 = -\lambda_3$ (HNC amplitudes). But HNC amplitudes are odd 
functions of $m$. So, we may expect that they should involve an additional 
power of $m$ as compared to the dominant amplitudes which are then 
HC amplitudes. On the other hand, if $n_\g$ is odd, 
$\Lambda+\lambda_p$ is even and then $\lambda_1 = \lambda_3$ (HC amplitudes). 
But HC amplitudes are even functions of $m$. So, we may expect that they 
should involve an additional power of $m$ as compared to the dominant 
amplitudes which are then HNC amplitudes.  

\vv \vv

\subsection{The subprocess 
$\gamma^{\star}(q)  \rightarrow \bar{\ell}(p_1) + \ell(p_3) + (N-1) 
{\rm{\bf real~photons}}(k_j)$} 

\vv \nin Let us now consider the process $\gamma^{\star}  \rightarrow \bar{\ell} + 
\ell + (N-1)~ \gamma's$ and let us try to derive, as before, some 
properties of its current when all final particles $\ell,~ \bar{\ell},~ 
(N-1)~ \gamma's$ are emitted in the strict forward direction.

\vv \nin Here, the generic form for numerators of helicity amplitudes is 

\begin{eqnarray}
&{\cal N} = \bar{U}_3 \g(\ep^{\star}_N)[m+\g(p_3+k_N)]\g(\ep^{\star}_{N-1})
[m+\g(p_3+k_N+k_{N-1})] \g(\ep^{\star}_{N-2})\cdots& \nonumber \\
& \times \g(\ep^{\star}_{p+1})
[m+\g(p_3+k_N+k_{N-1}+\cdots+k_{p+1})]~ \g_\mu ~ & \nonumber \\
&\times [m-\g(p_1+k_1+k_2+\cdots+k_{p-1})] \g(\ep^{\star}_{p-1}) &  \label{FW40}\\
&\times [m-\g(p_1+k_1+k_2-\cdots +k_{p-2})]\g(\ep^{\star}_{p-2})
\cdots[m-\g(p_1+k_1)]\g(\ep^{\star}_1)  V_{1c}& \nonumber
\end{eqnarray}

\nin where the vertex of the virtual photon stands at the $p$-th place (here too, 
we may set $q = - k_p$). The conjugate spinor $ V_{1c}$ is defined as
\beq
V_{1c}(\lambda_1) = - 2 \lambda_1 \g_5 U_1(-\lambda_1)
\label{FW41} \enq

\nin Of course, the above structure is quite analogous to that studied in the preceding subsection, 
simply because it can be derived from the latter by crossing. So, we will 
apply to (1) the same reductions as those used in subsection (4.1), and we will also 
use similar notations.
\bigskip

\noindent {\bf 1) \und{$n_{\g}$ is odd.}}\\
\bigskip

\noindent {\bf a) $n^a_{\g}$ even and $n^b_{\g}$ odd}
\begin{eqnarray}
&(- 2 \cp_3\cdot \ck_N)(- 2 \cp_3\cdot (\ck_N+\ck_{N-1}+\ck_{N-2}))\cdots 
(- 2 \cp_3\cdot (\ck_N+\ck_{N-1}+& \nonumber \\
& +\ck_{N-2}+\cdots+\ck_{p+2}))~ \bar{ \uz}_3 
~\g(E^{\star}_N)\g(E^{\star}_{N-1})
\g(E^{\star}_{N-2})\cdots \g(E^{\star}_{p+1})~\g_\mu~ &  \label{FW42}\\
& \g(\ck_1+\ck_2+\cdots+\ck_{p-1}) \g(E^{\star}_{p-1})\g(E^{\star}_{p-2})
\cdots \g(E^{\star}_1) \vz_{1c} & \nonumber  \\
& \times (-) (-2 \cp_1\cdot \ck_1)(-2 \cp_1\cdot (\ck_1+\ck_2+\ck_3))\cdots 
(-2 \cp_1\cdot (\ck_1+\ck_2+\cdots+\ck_{p-3})) & \nonumber
\end{eqnarray}

\noindent {\bf b) $n^a_{\g}$ odd and  $n^b_{\g}$ even}
\begin{eqnarray}
& (- 2 \cp_3\cdot \ck_N)
(- 2 \cp_3\cdot (\ck_N+\ck_{N-1}+\ck_{N-2}))\cdots 
(- 2 \cp_3\cdot (\ck_N+\ck_{N-1}+& \nonumber \\
&+\ck_{N-2}+\cdots+\ck_{p+3}))~\bar{ \uz}_3 ~
\g(E^{\star}_N)\g(E^{\star}_{N-1})
\g(E^{\star}_{N-2})\cdots \g(E^{\star}_{p+1}) & \nonumber \\
&\times  \g(\ck_N+\ck_{N-1}+\cdots+\ck_{p+1})~ \g_\mu~\g(E^{\star}_{p-1})
\g(E^{\star}_{p-2})\cdots \g(E^{\star}_1)  \vz_{1c} &  \label{FW43} \\
& \times (-2 \cp_1\cdot \ck_1)(-2 \cp_1\cdot (\ck_1+
\ck_2+\ck_3))\cdots 
(-2 \cp_1\cdot (\ck_1+\ck_2+\cdots+\ck_{p-2})) & \nonumber
\end{eqnarray} 
\bigskip
\biz 
\item From these two expression it is clear that when $n_{\g}$ 
is odd, all corresponding 
components $J_{-} = J\cdot(T+Z)$ 
of the current are strictly zero in the forward configuration.

\item Regarding impact factors $J_{+}=J\cdot (T-Z)$, they are non-zero only if 
\begin{eqnarray}
& \lambda_{\g_1} = -2 \lambda_1 = -\lambda_{\g_2}= \cdots = 
\lambda_{\g_{p-1}}= -\lambda_{\g_{p+1}}= & \nonumber \\
& \cdots=- \lambda_{\g_{N-1}} = \lambda_{\g_N} = - 2 \lambda_3~~~~
{\rm case~(a),~~~or}&  \label{FW44} \\
& \lambda_{\g_1} = -2 \lambda_1= -\lambda_{\g_2}= \cdots = 
- \lambda_{\g_{p-1}}=  \lambda_{\g_{p+1}}= & \nonumber \\
&\cdots = - \lambda_{\g_{N-1}} =\lambda_{\g_N} = - 2 \lambda_3~~~~{\rm case~(b)}& \nonumber 
\end{eqnarray}  

i.e. only for HC amplitudes. In this case, we find an amplitude proportional 
to the matrix element 
\beq
\bar{ \uz}_3 (\lambda_1)
\g(T \mp Z)\g(T \pm Z)\g(E^{(-2\lambda_1)\star})
\vz_{1c}(\lambda_1) 
\label{FW45} \enq

which is itself proportional to the following one  
\beq
\bar{ \uz}_3 (\lambda_1)
(1 \mp 2 \lambda_1 \g_5)\uz_1(\lambda_1) 
\label{FW46} \enq

and the latter is proportional to the lepton mass $m$.

Thus, we here find that only HC impact factors survive and should be 
proportional to 
\beq
(m^2)^{(N-p)/2} (m^2)^{(p-2)/2} m = m^{n_\g}
\label{FW47} \enq

\item The transverse components which are projections onto 
$E^{(\pm)} \equiv E^{(\lambda_{\g_p}) \star} $, are a priopri non-zero only if 
\begin{eqnarray}
& \lambda_{\g_1} = -2 \lambda_1 = -\lambda_{\g_2}= \cdots = 
\lambda_{\g_{p-1}}=-\lambda_{\g_{p}}= \lambda_{\g_{p+1}}= & \nonumber \\
& \cdots= \lambda_{\g_{N-1}} = - \lambda_{\g_N} =  2 \lambda_3~~~~
{\rm case~(a),~~~or}&  \label{FW48} \\
& \lambda_{\g_1} = -2 \lambda_1= -\lambda_{\g_2}= \cdots = 
- \lambda_{\g_{p-1}}=  \lambda_{\g_{p}}=- \lambda_{\g_{p+1}}= & \nonumber \\
&\cdots =  \lambda_{\g_{N-1}} =-\lambda_{\g_N} =  2 \lambda_3~~~~
{\rm case~(b)}& \nonumber 
\end{eqnarray}  

In this case, since we have now an even number of $\g( E^{\star})$, 
the amplitude is finally proportional to the matrix element 
\beq
\bar{ \uz}_3(\lambda_1) \g(T+Z) \uz_1(\lambda_1) 
\label{FW49} \enq

which is itself proportional to $m^2$. Thus, only HNC transverse components 
survive and they are proportional to \und{$m^{n_\g+1}$}.
\eiz

\vv \nin It is worth noticing that ``mass-parity'' seems here different to that in 
subsection (4.1). But this is simply due to the choice we made to define the 
helicity of the 
antilepton. With the definition (\ref{FW41}), amplitudes are proportional to the phase 
factor $\exp(i\lambda_1 \varphi_1)$ (as if we had an outgoing lepton), and the 
general rule $\Lambda +\lambda_3 + \lambda_1 = n -p$ should be applied. 
Taking $n=p=0$, we get, when $n_{\g}$ is odd : 

\biz
\item $\lambda_3= \lambda_1$ for impact factor, because then $\Lambda$, 
which is the sum of the helicities of the outgoing photons, is 
odd ; hence, HC amplitudes ;  
\item $\lambda_3= - \lambda_1$ for transverse components, because 
$\Lambda$, which should include the $\pm 1$ helicity of the virtual photon, 
is now even ; hence, HNC amplitudes.
\eiz  

\vv \nin  {\bf 2) \und{$n_{\g}$ is even.}}\\

\vv \nin  {\bf a) $n^a_{\g}$ and $n^b_{\g}$ are even}
\begin{eqnarray}
&(- 2 \cp_3\cdot \ck_N)(- 2 \cp_3\cdot (\ck_N+\ck_{N-1}+\ck_{N-2}))\cdots 
(- 2 \cp_3\cdot (\ck_N+\ck_{N-1}+& \nonumber \\
& +\ck_{N-2}+\cdots+\ck_{p+2}))~ \bar{ \uz}_3 
~\g(E^{\star}_N)\g(E^{\star}_{N-1})
\g(E^{\star}_{N-2})\cdots \g(E^{\star}_{p+1})~\g_\mu~ &  \label{FW50} \\
& \g(E^{\star}_{p-1})\g(E^{\star}_{p-2})
\cdots \g(E^{\star}_1) \vz_{1c} & \nonumber  \\
& \times (-2 \cp_1\cdot \ck_1)(-2 \cp_1\cdot (\ck_1+\ck_2+\ck_3))\cdots 
(-2 \cp_1\cdot (\ck_1+\ck_2+\cdots+\ck_{p-2})) & \nonumber
\end{eqnarray}

\noindent {\bf b) $n^a_{\g}$ and  $n^b_{\g}$ are odd}
\begin{eqnarray}
& (- 2 \cp_3\cdot \ck_N)
(- 2 \cp_3\cdot (\ck_N+\ck_{N-1}+\ck_{N-2}))\cdots 
(- 2 \cp_3\cdot (\ck_N+\ck_{N-1}+& \nonumber \\
&+\ck_{N-2}+\cdots+\ck_{p+3}))~\bar{ \uz}_3 ~
\g(E^{\star}_N)\g(E^{\star}_{N-1})
\g(E^{\star}_{N-2})\cdots \g(E^{\star}_{p+1}) & \nonumber \\
&\times  \g(\ck_N+\ck_{N-1}+\cdots+\ck_{p+1})~ \g_\mu~
 \g(\ck_1+\ck_2+\cdots+\ck_{p-1}) &  \label{FW51} \\
&\times \g(E^{\star}_{p-1})
\g(E^{\star}_{p-2})\cdots \g(E^{\star}_1)  \vz_{1c}\times(-) 
(-2 \cp_1\cdot \ck_1)  & \nonumber \\
& \times (-2 \cp_1\cdot (\ck_1+
\ck_2+\ck_3))\cdots 
(-2 \cp_1\cdot (\ck_1+\ck_2+\cdots+\ck_{p-3})) & \nonumber
\end{eqnarray} 
\bigskip
\biz
\item $J_{-}$ components exist only in case a). We must have  
\begin{eqnarray}
&\lambda_{\g_1} = -2 \lambda_1 = -\lambda_{\g_2}= \cdots = 
- \lambda_{\g_{p-1}} = & \label{FW52} \\
&\lambda_{\g_{p+1}}= \cdots= \lambda_{\g_{N-1}} 
= - \lambda_{\g_N} =  2 \lambda_3 & \nonumber
\end{eqnarray}  

i.e. HNC amplitudes (in agreement with the general rule) 
that are proportional to the matrix element 
\beq
\bar{ \uz}_3(-\lambda_1) \g(T+Z) \g_5 \uz_1
(-\lambda_1)
\label{FW53} \enq

which is proportional to $m^2$. Thus, in this case HNC $J_{-}$'s are 
\und{$\propto m^{n_\g + 2}$}.

\item In both cases, impact factors are also found as HNC amplitudes (even 
number of $\g(E^{\star})$'s). In case a) they are proportional to   
\beq
\bar{ \uz}_3(-\lambda_1) \g(T-Z) \g_5 \uz_1
(-\lambda_1)
\label{FW54} \enq

which is of order zero in $m$, while in case b) they are proportional to 
\beq
\bar{ \uz}_3(-\lambda_1) \g(T+Z) \g_5 \uz_1
(-\lambda_1)
\label{FW55} \enq

which is $\propto m^2$. Thus, in both cases, we find HNC impact factors 
\und{$\propto m^{n_{\g}}$}.

\item Finally, transverse components are HC amplitudes and it is not difficult 
to show that they are zero in case b) and \und{$\propto m^{n_{\g}+1}$} in 
case a), being $\propto$ 
\beq
\bar{\uz}_3(\lambda_1) \g(E^{(-2\lambda_1) \star}) 
\vz_{1c}(\lambda_1) \sim
\bar{\uz}_3(\lambda_1) \uz_1 (\lambda_1) \propto m
\label{FW56} \enq
\eiz

\section{The case of multi-bremsstrahlung processes} 
 
\subsection{ Introduction}

\vv \nin Let us remind that the name ``order'' here refers to jet-like kinematics,
where all particles are ultra-relativistic (for massive ones their energies $E$ are
$\gg m$), and all final particles of a given vertex are emitted near the
direction of propagation of the parent incoming particle. Terms of order
one are $m/E$ and polar angles $\theta$. However, in the following, by ``order'' of a term or
of an expression we mean their ``lowest order'' in the jet-like
kinematics approximation. Thus, $\cos(\theta)$ is said to be of order zero and $\sin(\theta)$ of order one ;
$p_{-}= E-p_Z$ is of order two, etc.

\vv \nin In dealing with approximation, we have to find an efficient
method providing a line of action for picking all necessary terms
that yield a given order of approximation, of course without forgetting none of
these terms. The problem we have in mind is the following.

\vv \nin Consider the generic form of an impact factor for the multi-bremsstrahlung process lepton +
$\gamma^\star \rightarrow$ lepton + $n_\gamma$ real photons with
$n_\gamma = N-1$ :

$$ J = \bar{U}_3 {\cal M} U_1 ~~~~{\rm with} $$
$$ {\cal M} = \g(\epsilon^\star_N) [ m + \g(p_3 + k_N) ]
\g(\epsilon^\star_{N-1}) [ m + \g(p_3 + k_N + k_{N-1})
]\g(\epsilon^\star_{N-2}) \cdots $$
$$ \times \g(\epsilon^\star_{p+1})[ m + \g(p_3 + k_N + k_{N-1}+\cdots + k_{p+1})
]\g(T-Z) $$
\beq \times [ m + \g(p_1 - k_1 - k_2 - \cdots -
k_{p-1})] \g(\epsilon^\star_{p-1}) \cdots  \label{AMB} \enq
$$ \times [ m + \g(p_1 - k_1 - k_2 - \cdots - k_{p-2})]
\g(\epsilon^\star_{p-2}) \cdots [ m + \g(p_1 - k_1)] \g(\epsilon_1)
$$

\vv \nin where the notation $\g(V) = V_\mu \gamma^\mu$ is used here again and propagators have been omitted. Notice that the impact
factor $J$ in (\ref{AMB}) has mass dimension $p-1 + N-p +1 = n_\g +1$ (spinors
being normalized according to $\bar{U} U = 2 m$). 

\vv \nin We will use the following decompositions.  For any
4-momentum
$Q$, we have

$$ \g(Q) = \di{Q_{+}\over 2} \g(T+Z) + \di{Q_{-}\over 2} \g(T-Z)
+ \g(Q_T)$$
\beq {\rm where}~~~
Q_T = Q_X X + Q_Y Y,~~~~Q_{\pm} = Q_0 \pm Q_z
\label{DEC1} \enq

\vv \nin and for any polarization 4-vector, we take (see (\ref{polar})) 

\beq
\g(\epsilon^{\star}) = \exp(i \Lambda \varphi_{\g})~[ \g(E^{
(\Lambda) \star})+ \xi~ \g(T-Z)]
\label{DEC2}
\enq

\vv \nin Since the leading order of  ${\cal M}$ in (\ref{AMB})
is $N-1 = n_\gamma$, we should find that all terms of less
order in its development are strictly zero ! This can be easily checked for the term of zeroth \mbox{order :} 
let us take the leading orders in (\ref{DEC1}) and (\ref{DEC2}) 

$$ \g(Q) \approx \di{Q_{+} \over 2} \g(T+Z) ~,~~~\g(\epsilon^{(\Lambda) \star} ) \approx \exp{(i \Lambda \varphi_\gamma)} \g(E^{(\Lambda) \star}) $$

\vv \nin then 

$${\cal M}^{(0)} \propto \g(E^\star_N) \g(T+Z) \g(E^\star_{N-1}) \g(T+Z) \cdots $$
\beq \times \g(E^\star_{p+1}) \g(T+Z) \g(T-Z) \g(T+Z) \g(E^\star_{p-1}) \cdots \g(T+Z) \g(E^\star_1 ) \enq

\vv \nin which is identically zero since $\g(E^\star_k)$ and $\g(T+Z)$ are anticommuting and $[\g(T+Z)]^2 =0$. So, to pick terms of order $n_\gamma$ is not so trivial as we are forced to make the development of the product of $N-1$ factors, losing then transparency in calculation : having $n_\gamma$ factors $F_k$ that we expand according various orders in the form 

$$ F_k = F^{(0)}_k + F^{(1)}_k + F^{(2)}_k + \cdots $$   

\vv \nin ${\cal M} = F_1 F_2 F_3 \cdots F_{n_\gamma}$ should have the leading form 

$$ {\cal M} \simeq \di{\sum_{n_1 + n_2 + \cdots n_{n_\gamma} = n_\gamma}} F^{(n_1)}_1 F^{(n_2)}_2 \cdots F^{(n_{n_\gamma)}}_{n_\gamma} $$

\vv \nin This shows that obtaining the lowest order expression of an impact factor is generally not an easy task, and, as in any computation of approximate forms, the legitime questions arise : how do the various terms combine so as to yield the lowest-order expression ? Is it possible to discard a priori some terms ? 

\vv \nin Our goal is then twofold. First, it is highly desirable to have a full
control over the orders of the various terms entering a given transition
amplitude, with the aim to make coherent the approximations one is led
to perform in the framework of jet-like kinematics. Secondly, we would like
to find some systematics for easy computation, if possible. In the following,
we attempt to get some clues to this problem, and to develop a formalism that
could be useful for computational purposes.

\subsection{Tracking orders in a computer's dream}

\vv \nin Let us consider a matrix of the form
\beq
{\cal N} = \left[m + \g(Q)\right]~\g(\epsilon^{\star})
\label{matrix1}
\enq

\vv \nin Using the above decompositions (\ref{DEC1}) and (\ref{DEC2}), we get
\begin{eqnarray}
&{\cal N} = \left[m + \di{Q_{+}\over 2} \g(T+Z) + \di{Q_{-}\over 2}
\g(T-Z) + \g(Q_T)\right]~\exp(i \Lambda \varphi)~\times&
\nonumber \\
&[ \g(E^{\star})+ \xi~ \g(T-Z)] = \di{Q_{+}\over 2}
\g(T+Z)\g(E^{\star}) + \left[m +\g(Q_T)\right]
\g(E^{\star}) + & \nonumber \\
&\xi~\di{Q_{+}\over 2} \g(T+Z)\g(T-Z) +\left[m +\g(Q_T)\right]
\xi~\g(T-Z) +
 & \\
&\di{Q_{-}\over 2}\g(T-Z)\g(E^{\star}) = -
\di{Q_{+}\over 2}\g(T+Z)\g(E^{\star})+\left[m +
\g(Q_T)\right]\g(E^{\star}) + & \nonumber \\
&\xi~\di{Q_{+}\over 2}
\g(T+Z)\g(T-Z) +
\left[m +\g(Q_T)\right]
\xi~\g(T-Z) -\di{Q_{-}\over 2}\g(E^{\star})
\g(T-Z)& \nonumber
\end{eqnarray}

\vv \nin which can be rewritten in the form (omitting the phase factor
$\exp(i \Lambda \varphi)$)

\beq
{\cal N} = a \g(T+Z) + b + c \g(T-Z)
\label{matrix2}
\enq

\nin where

\begin{eqnarray}
&a = - \di{Q_{+}\over 2}\g(E^{\star})& \nonumber \\
\label{matrix3}
&b = \left[m + \g(Q_T)\right]\g(E^{\star}) +\xi~\di{Q_{+}\over 2}
\g(T+Z)\g(T-Z) &  \\
&c =\xi~ \left[m +\g(Q_T)\right]-\di{Q_{-}\over 2}
\g(E^{\star})& \nonumber
\end{eqnarray}

\vv \nin The above arrangement has been performed according to the lowest orders
of the various terms entering
the matrix ${\cal N}$ : thus, $a$ is of (lowest)
order $0$, $b$ is of (lowest) order $1$ and
$c$ is of (lowest) order $2$. In the following, by ``order'' of a term or
of an expression we will mean their ``lowest order'' in the jet-like
kinematics approximation. Thus, $\theta$ being a polar
angle, $\cos(\theta)$ is of order zero and $\sin(\theta)$ is of order one ;
$p_{-}= E-p_Z$ is of order two, etc.

\vv \nin The matrices $a$, $b$ and $c$ have the following properties

\begin{eqnarray}
&a \g(T \pm Z)  = \g(T \pm Z) a^{(1)}~~~{\rm with}~~~ a^{(1)} =
-a & \label{fund1}
\nonumber \\
& \g(T \pm Z) a  = a^{(1)} \g(T \pm Z) & \nonumber \\
& \g(T+Z) b =  b^{(1)}\g(T+Z)~~~{\rm with}~~~ b^{(1)} = \left[\g(Q_T) -
m\right]\g(E^{\star}) & \nonumber \\
& \g(T-Z) b  = b^{(2)} \g(T-Z)~~~{\rm with}~~~ b^{(2)} = b^{(1)} +
\xi~\di{Q_{+}\over 2}\g(T-Z)\g(T+Z) & \nonumber \\
&b \g(T+Z)  =  \g(T+Z) b^{(2)}  &  \\
& b \g(T-Z)  = \g(T-Z) b^{(1)}& \nonumber \\
& \g(T \pm Z) b^{(1)}   = b^{(3)} \g(T \pm Z)~~~{\rm with}~~~ b^{(3)} =
\left[m +
\g(Q_T)\right]\g(E^{\star}) & \nonumber \\
& b^{(1)} \g(T \pm Z) =  \g(T \pm Z) b^{(3)} \nonumber \\
& c \g(T \pm Z) = \g(T \pm Z) c^{(1)}~~~{\rm with}~~~c^{(1)} = \left[m -
\g(Q_T)\right] \xi + \di{Q_{-}\over 2}\g(E^{\star})& \nonumber \\
& \g(T \pm Z) c  =  c^{(1)}\g(T \pm Z)  & \nonumber
\end{eqnarray}

\vv \nin Let us stress that in the expression (\ref{matrix2}) not only
terms of various orders are well separated, but also matrices
$\g(T+Z)$, $1$ and $\g(T-Z)$ factorize, respectively, the zeroth
order term, the first order term and the second order term.
This suggests some simple underlying algebra.

\vv \nin Let us thus define matrices
\beq
{\cal N}_i = p_i + x q_i + x^2 r_i
\label{matrix4}
\enq

\vv \nin where $x$ is a real parameter and where
\beq
p_i = a_i \g(T+Z),~~~q_i = b_i,~~~r_i = c_i\g(T-Z)
\enq

\vv \nin the matrix factors $a_i$, $b_i$ and $c_i$ being defined as in
(\ref{matrix3}) with appropriate labels. As will be seen below, introducing the parameter
$x$ provides a useful tool to track orders. The previous forms
(\ref{matrix2}) of matrices
${\cal N}_i$ can be recovered by taking $x = 1$. Below we quote again
the fundamental properties
of matrices $a_i$, $b_i$ and $c_i$ that will be important in the following :

\begin{eqnarray}
& \g(T\pm Z) a_i  = a^{(1)}_i \g(T \pm Z),~~~a_i \g(T\pm Z) =
\g(T \pm Z) a^{(1)}_i  & \label{fund2} \nonumber \\
& \g(T+Z) b_i =  b^{(1)}_i \g(T+Z),~~~b_i \g(T-Z) = \g(T-Z) b^{(1)}_i
&  \\
&  b_i \g(T+Z) = \g(T+Z) b^{(2)}_i,~~~\g(T-Z) b_i = b^{(2)}_i \g(T-Z)
& \nonumber \\
& \g(T\pm Z) c_i  = c^{(1)}_i \g(T \pm Z),~~~c_i \g(T\pm Z) =
\g(T \pm Z) c^{(1)}_i  & \nonumber
\end{eqnarray}

\vv \nin Let us consider the product of two matrices ${\cal N}_1$ and  ${\cal N}_2$.
Since $\g(T\pm Z)^2 = 0$, we have

\begin{eqnarray}
&{\cal N}_2{\cal N}_1 = \left(a_2 \g(T+Z) + b_2 x+ x^2c_2 \g(T-Z)\right)
\left( \g(T+Z)a^{(1)}_1 + x b_1 + \right.
& \nonumber \\
& \left. x^2 \g(T-Z)c^{(1)}_1\right) =  x a_2 \g(T+Z) b_1 + x^2 a_2
\g(T+Z)\g(T-Z)c^{(1)}_1 + & \\
&  x b_2
\g(T+Z)a^{(1)}_1
+ x ^2 b_2 b_1 +x^3 b_2\g(T-Z)c^{(1)}_1 +
& \nonumber \\
& x^2 c_2 \g(T-Z)\g(T+Z)a^{(1)}_1 +  x^3 c_2
\g(T-Z) b_1 & \nonumber
\end{eqnarray}

\vv \nin Using the above-mentioned properties (\ref{fund2}) of matrices
$a$, $b$ and $c$, it is
possible to rewrite this product in the same form as (21) :

\beq
{\cal N}_2{\cal N}_1 =x \left( A_2 \g(T+Z) +x B_2 +x^2 C_2 \g(T-Z) \right)
\enq
with
\begin{eqnarray}
&A_2 = b_2 a_1 + a_2 b^{(1)}_1 & \nonumber \\
&B_2 = b_2 b_1 + c_2 a^{(1)}_1 \g(T-Z)\g(T+Z) + a_2 c^{(1)}_1
\g(T+Z)\g(T-Z)& \\
&C_2 = c_2  b^{(2)}_1 + b_2 c_1 & \nonumber
\end{eqnarray}

\vv \nin As is revealed by the overall factor $x$, the multiplication of the two
matrices increases orders by one unit. We see also that the above generic
form provides again a clean separation of terms having different
orders. Moreover, the hierarchy of orders is \mbox{preserved :} $A_2$ is
now of first order, $B_2$ is of second
order and $C_2$ of third order.

\vv \nin The new matrices $A_2$, $B_2$ and $C_2$ have the following properties :

\begin{eqnarray}
&A_2 \g(T+Z)  = \g(T+Z) A^{(1)}_2,~~~~\g(T-Z) A_2 = A^{(1)}_2\g(T-Z)
& \label{FUN1}\nonumber \\
&A_2 \g(T-Z)  = \g(T-Z) A^{(2)}_2,~~~~\g(T+Z)A_2  =  A^{(2)}_2\g(T+Z)&  \\
& A^{(1)}_2\g(T+Z) = \g(T+Z)A^{(3)}_2,~~~~\g(T-Z) A^{(1)}_2=A^{(3)}_2\g(T-Z)&
\nonumber
\end{eqnarray}

\nin with

\begin{eqnarray}
&A^{(1)}_2 = b^{(2)}_2 a^{(1)}_1 + a^{(1)}_2 b^{(3)}_1,~A^{(2)}_2 =
b^{(1)}_2a^{(1)}_1 + a^{(1)}_2  b^{(3)}_1,~A^{(3)}_2 =b^{(3)}_2 a_1 +
a_2 b^{(1)}_1 &
\end{eqnarray}

\begin{eqnarray}
& \g(T+Z) B_2 =  B^{(1)}_2\g(T+Z),~~~~B_2 \g(T-Z) = \g(T-Z)B^{(1)}_2
& \label{FUN2}\nonumber \\
& \g(T-Z) B_2  = B^{(2)}_2 \g(T-Z),~~~~ B_2 \g(T+Z)  = \g(T+Z)B^{(2)}_2
&  \\
&B^{(1)}_2\g(T-Z)=\g(T-Z)B^{(3)}_2,~~~~\g(T+Z)B^{(1)}_2=B^{(3)}_2\g(T+Z)
& \nonumber
\end{eqnarray}

\nin with

\begin{eqnarray}
&B^{(1)}_2 =  b^{(1)}_2 b^{(1)}_1 +
\g(T+Z) c_2 a^{(1)}_1\g(T-Z) + \g(T-Z)c_2 a^{(1)}_1\g(T+Z) & \nonumber \\
&B^{(2)}_2 =  b^{(2)}_2 b^{(2)}_1
+\g(T-Z) a_2 c^{(1)}_1\g(T+Z)+\g(T+Z) a_2 c^{(1)}_1\g(T-Z) & \\
& B^{(3)}_2 = b^{(3)}_2 b^{(3)}_1 +\g(T+Z)c^{(1)}_2 a_1 \g(T-Z)+
\g(T-Z)c^{(1)}_2 a_1 \g(T+Z) & \nonumber
\end{eqnarray}

\begin{eqnarray}
& C_2 \g(T-Z) = \g(T-Z) C^{(1)}_2,~~~\g(T+Z) C_2 =  C^{(1)}_2\g(T+Z)&
\label{FUN3}\nonumber \\
& C_2 \g(T+Z) = \g(T+Z) C^{(2)}_2,~~~\g(T-Z) C_2 =  C^{(2)}_2\g(T-Z) &
\end{eqnarray}
with
\begin{eqnarray}
&C^{(1)}_2 = c^{(1)}_2 b_1+b^{(1)}_2 c^{(1)}_1,~~C^{(2)}_2 = c^{(1)}_2
b^{(3)}_1 +b^{(2)}_2 c^{(1)}_1 &
\end{eqnarray}

\vv \nin Thus, it is seen that due to the structure of the matrices
$A$, $B$ or $C$, which are constructed from $\g(X)$, $\g(Y)$, and
from the products $\g(T + Z)\g(T - Z)$ and $\g(T- Z)\g(T + Z)$,
any product like $\g(T\pm Z) {\cal X}$ or
${\cal X} \g(T\pm Z)$, ${\cal X}$ being $A$, $B$ or $C$,
gives ${\cal X}'\g(T\pm Z)$ and $\g(T\pm Z){\cal X}''$ respectively,
where ${\cal X}'$ and ${\cal X}''$ have a similar structure as that
of ${\cal X}$. As a consequence, the product of any number of matrices
such as ${\cal N}$ has the same structure as that of ${\cal N}$. To prove
this explicitly, we may proceed by recurrence. Let $N_n$ be the
matrix $N_n =
A_n \g(T+Z) + B_n + C_n \g(T-Z)$ where the matrices $A_n$, $B_n$ and $C_n$
have the general properties described by formulas (\ref{FUN1}),
(\ref{FUN2}) and (\ref{FUN3}),
and ${\cal N}_{n+1}$ a matrix like (\ref{matrix1}), (\ref{matrix2}). We have
\begin{eqnarray}
&N_{n+1}= {\cal N}_{n+1} N_n = \left[a_{n+1} \g(T+Z) + b_{n+1} + c_{n+1}
\g(T-Z)\right]
\left[A_n \g(T+Z) +\right.  & \nonumber \\
& \left. + B_n + C_n \g(T-Z)\right] = a_{n+1} \g(T+Z) A_n \g(T+Z) +
a_{n+1} \g(T+Z) B_n + & \nonumber\\
& a_{n+1} \g(T+Z) C_n\g(T-Z) + b_{n+1}  A_n \g(T+Z) + b_{n+1} B_n +
 &  \nonumber \\
& b_{n+1} C_n \g(T-Z) + c_{n+1} \g(T-Z) A_n \g(T+Z) + c_{n+1} \g(T-Z) B_n +
 & \\
&c_{n+1} \g(T-Z) C_n \g(T-Z) = a_{n+1} A^{(1)}_n \g(T+Z)^2 +
a_{n+1} B^{(1)}_n \g(T+Z) +  & \nonumber \\
&  \g(T+Z)
a^{(1)}_{n+1} C_n \g(T-Z) +b_{n+1}  A_n \g(T+Z) + b_{n+1} B_n + b_{n+1}
C_n \g(T-Z) + &  \nonumber \\
& \g(T-Z)c^{(1)}_{n+1} A_n \g(T+Z) + c_{n+1} B^{(2)}_n \g(T-Z) +
c_{n+1} \g(T-Z)^2 C^{(1)}_n & \nonumber
\end{eqnarray}

\vv \nin Taking care of hierarchy of orders, this product may be written in the form
\beq
N_{n+1}=A_{n+1} \g(T+Z) + B_{n+1} + C_{n+1} \g(T-Z)
\enq

\nin where

\begin{eqnarray}
& A_{n+1} =  b_{n+1}  A_n + a_{n+1} B^{(1)}_n  & \label{REC1} \nonumber \\
& B_{n+1} =  b_{n+1} B_n + \g(T-Z)c^{(1)}_{n+1} A_n \g(T+Z) +
\g(T+Z)a^{(1)}_{n+1} C_n\g(T-Z) &   \\
& C_{n+1} =   c_{n+1} B^{(2)}_n +  b_{n+1} C_n &  \nonumber
\end{eqnarray}

\vv \nin In fact, the preceding calculation shows that it is sufficient to prove that
the relations

\begin{eqnarray}
&A_n \g(T+Z)  = \g(T+Z) A^{(1)}_n & \label{SUF} \nonumber \\
& \g(T+Z) B_n =  B^{(1)}_n\g(T+Z)& \nonumber \\
& \g(T- Z)B_n  = B^{(2)}_n \g(T-Z) &  \\
&  C_n \g(T-Z)   = \g(T-Z)C^{(1)}_n  & \nonumber
\end{eqnarray}

\vv \nin hold true for any matrix ${\cal N}$, i.e. for all $n$. They are obviously true for $n = 1$.
Assuming they are true for some $n$, i.e. for
the matrices $A_n$, $B_n$ and $C_n$, then, from the recurrence relations (\ref{REC1}) we have

\begin{eqnarray}
&A_{n+1} \g(T+Z)  =  a_{n+1} B^{(1)}_n\g(T+Z) +  b_{n+1}  A_n \g(T+Z) =  &
\nonumber \\
& \g(T+Z)a^{(1)}_{n+1} B_n + \g(T+Z)b^{(2)}_{n+1} A^{(1)}_n = &  \\
&\g(T+Z)A^{(1)}_{n+1}~~~~{\rm with}~~~~A^{(1)}_{n+1}= a^{(1)}_{n+1} B_n +
b^{(2)}_{n+1} A^{(1)}_n&  \nonumber
\end{eqnarray}

\begin{eqnarray}
& \g(T+Z) B_{n+1} = \g(T+Z) b_{n+1} B_n + \g(T+Z)\g(T-Z)c^{(1)}_{n+1} A_n
\g(T+Z) = & \nonumber \\
& b^{(1)}_{n+1} B^{(1)}_n\g(T+Z) + \g(T+Z) \g(T-Z) c^{(1)}_{n+1} A_n \g(T+Z) =
& \\
& B^{(1)}_{n+1}\g(T+Z)~~~~{\rm with}~~~~~B^{(1)}_{n+1} = b^{(1)}_{n+1}
B^{(1)}_n +
\g(T+Z) \g(T-Z) c^{(1)}_{n+1} A_n & \nonumber
\end{eqnarray}

\begin{eqnarray}
& \g(T- Z) B_{n+1} = \g(T-Z) b_{n+1} B_n + \g(T-Z)
\g(T+Z)a^{(1)}_{n+1} C_n \g(T-Z)= & \nonumber \\
& b^{(2)}_{n+1} B^{(2)}_n \g(T-Z) + \g(T-Z)
\g(T+Z)a^{(1)}_{n+1} C_n \g(T-Z)= & \\
& B^{(2)}_{n+1}\g(T-Z)~~~~{\rm with}~~~~~B^{(2)}_{n+1} = b^{(2)}_{n+1}
B^{(2)}_n +
\g(T-Z) \g(T+Z)a^{(1)}_{n+1} C_n & \nonumber
\end{eqnarray}

\begin{eqnarray}
& C_{n+1} \g(T- Z) = c_{n+1} B^{(2)}_n \g(T-Z)  + b_{n+1} C_n
\g(T-Z) = & \nonumber \\
& \g(T-Z) c^{(1)}_{n+1} B_n +\g(T-Z) b^{(1)}_{n+1} C^{(1)}_n =\g(T-Z)
C^{(1)}_{n+1} &  \\
& {\rm with}~~~~~C^{(1)}_{n+1} = c^{(1)}_{n+1} B_n +
b^{(1)}_{n+1} C^{(1)}_n & \nonumber
\end{eqnarray}

\vv \nin Thus, properties (\ref{SUF}) also hold for the matrices $A_{n+1}$,
$B_{n+1}$ and
$C_{n+1}$ ; Q.E.D.

\vv \nin Now, if all properties given in (\ref{FUN1}), (\ref{FUN2}) and
(\ref{FUN3}) hold for the matrices
$A_n$, $B_n$ and $C_n$, then we have in addition

\begin{eqnarray}
& \g(T-Z) A_{n+1}  = \g(T-Z) a_{n+1} B^{(1)}_n + \g(T-Z) b_{n+1} A_n =
& \nonumber \\
& a^{(1)}_{n+1} B_n \g(T-Z) + b^{(2)}_2 A^{(1)}_n\g(T-Z)
=A^{(1)}_{n+1}\g(T-Z) &
\end{eqnarray}
\begin{eqnarray}
&A_{n+1} \g(T-Z) =  a_{n+1} B^{(1)}_n\g(T-Z) +  b_{n+1}  A_n \g(T-Z) =
& \nonumber \\
& \g(T-Z) a^{(1)}_{n+1} B^{(3)}_n + \g(T-Z)b^{(1)}_{n+1} A^{(2)}_n =
\g(T-Z)A^{(2)}_{n+1} &  \\
&{\rm with}~~~~A^{(2)}_{n+1}= a^{(1)}_{n+1} B^{(3)}_n +
b^{(1)}_{n+1} A^{(2)}_n &  \nonumber
\end{eqnarray}
\begin{eqnarray}
& \g(T+Z) A_{n+1}  = \g(T+Z) a_{n+1} B^{(1)}_n + \g(T+Z) b_{n+1} A_n
= & \nonumber \\
& a^{(1)}_{n+1} B^{(3)}_n \g(T+Z) + b^{(1)}_{n+1} A^{(2)}_n \g(T+Z)
=A^{(2)}_{n+1} \g(T+Z) &
\end{eqnarray}
\begin{eqnarray}
& A^{(1)}_{n+1} \g(T+Z) = a^{(1)}_{n+1} B_n \g(T+Z) + b^{(2)}_{n+1} A^{(1)}_n
\g(T+Z) = & \nonumber \\
& \g(T+Z)a_{n+1} B^{(2)}_n +\g(T+Z)b^{(3)}_{n+1} A^{(3)}_n
=\g(T+Z)A^{(3)}_{n+1} &  \\
&{\rm with}~~~~A^{(3)}=  a_{n+1} B^{(2)}_n + b^{(3)}_{n+1} A^{(3)}_n
&  \nonumber
\end{eqnarray}
\begin{eqnarray}
& \g(T-Z)A^{(1)}_{n+1} = \g(T-Z)a^{(1)}_{n+1} B_n +\g(T-Z) b^{(2)}_{n+1}
A^{(1)}_n = & \nonumber \\
& a_{n+1} B^{(2)}_n \g(T-Z) + b^{(3)}_{n+1} A^{(3)}_n \g(T-Z)
=A^{(3)}_{n+1} \g(T+Z) &
\end{eqnarray}

\begin{eqnarray}
&\g(T+Z)B_{n+1} = \g(T+Z)b_{n+1} B_n + \g(T+Z) \g(T-Z)c^{(1)}_{n+1}
A_n \g(T+Z) = & \nonumber \\
& b^{(1)}_{n+1} B^{(1)}_n \g(T+Z) +
\g(T+Z)c_{n+1} A^{(1)}_n \g(T-Z)\g(T+Z) &  \\
& =\g(T+Z)B^{(1)}_{n+1} & \nonumber
\end{eqnarray}
with
\begin{eqnarray}
&B^{(1)}_{n+1} = b^{(1)}_{n+1} B^{(1)}_n +
\g(T+Z)c_{n+1} A^{(1)}_n \g(T-Z) +  & \nonumber  \\
& \g(T-Z)c_{n+1} A^{(1)}_n \g(T+Z) &
\end{eqnarray}

\begin{eqnarray}
&B_{n+1} \g(T-Z)   = b_{n+1} B_n \g(T-Z) + \g(T-Z)c^{(1)}_{n+1} A_n
\g(T+Z)\g(T-Z) = & \nonumber \\
&\g(T-Z) b^{(1)}_{n+1} B^{(1)}_n  + \g(T-Z)\g(T+Z) c_{n+1} A^{(1)}_n
\g(T-Z) = &   \\
& \g(T-Z)B^{(1)}_{n+1} &  \nonumber
\end{eqnarray}

\begin{eqnarray}
& \g(T-Z)B_{n+1}  =\g(T-Z) b_{n+1} B_n  + \g(T-Z)\g(T+Z)a^{(1)}_{n+1} C_n
\g(T-Z) = & \nonumber \\
&b^{(2)}_{n+1} B^{(2)}_n \g(T-Z) + \g(T-Z)a_{n+1} C^{(1)}_n \g(T+Z)\g(T-Z)
= &   \\
& B^{(2)}_{n+1} \g(T-Z) &  \nonumber
\end{eqnarray}
with
\begin{eqnarray}
&B^{(2)}_{n+1} = b^{(2)}_{n+1} B^{(2)}_n + \g(T-Z)a_{n+1} C^{(1)}_n \g(T+Z)
+  &  \nonumber \\
&\g(T+Z)a_{n+1} C^{(1)}_n \g(T-Z) &
\end{eqnarray}
\begin{eqnarray}
& B_{n+1} \g(T+Z)  = b_{n+1} B_n \g(T+Z)  + \g(T+Z)a^{(1)}_{n+1} C_n \g(T-Z)
\g(T+Z) = & \nonumber \\
&\g(T+Z)b^{(2)}_{n+1} B^{(2)}_n  + \g(T+Z)a_{n+1} C^{(1)}_n \g(T-Z)\g(T+Z)
= &   \\
&\g(T+Z) B^{(2)}_{n+1} &  \nonumber
\end{eqnarray}

\begin{eqnarray}
& B^{(1)}_{n+1} \g(T-Z)  = b^{(1)}_{n+1} B^{(1)}_n \g(T-Z) +
\g(T-Z)\g(T+Z)c_{n+1} A^{(1)}_n \g(T-Z) = &\nonumber  \\
& \g(T-Z) b^{(3)}_{n+1} B^{(3)}_n  + \g(T-Z)\g(T+Z)c^{(1)}_{n+1}
A^{(3)}_n \g(T-Z)=  &  \nonumber \\
& \g(T-Z)B^{(3)}_{n+1} &
\end{eqnarray}
with
\begin{eqnarray}
&B^{(3)}_{n+1} = b^{(3)}_{n+1} B^{(3)}_n + \g(T+Z)c^{(1)}_{n+1}
A^{(3)}_n \g(T-Z)+   & \\
&\g(T-Z) c^{(1)}_{n+1}
A^{(3)}_n \g(T+Z) & \nonumber
\end{eqnarray}

\begin{eqnarray}
& \g(T+Z) B^{(1)}_{n+1}  = \g(T+Z)b^{(1)}_{n+1} B^{(1)}_n +
\g(T+Z)\g(T-Z)c_{n+1} A^{(1)}_n \g(T+Z) = &\nonumber  \\
& b^{(3)}_{n+1} B^{(3)}_n \g(T+Z) + \g(T+Z)c^{(1)}_{n+1}
A^{(3)}_n \g(T-Z)\g(T+Z) =  &  \\
& B^{(3)}_{n+1} \g(T+Z) &
\end{eqnarray}

\begin{eqnarray}
&\g(T+Z) C_{n+1}   =\g(T+Z) c_{n+1} B^{(2)}_n +  \g(T+Z) b_{n+1} C_n
= & \nonumber \\
& c^{(1)}_{n+1} B_n \g(T+Z) + b^{(1)}_{n+1} C^{(1)}_{n+1}\g(T+Z) =
C^{(1)}_{n+1} \g(T+Z) &
\end{eqnarray}

\vv \nin Without prejudging the practical usefulness of the above
development, at least the latter has the advantage of providing us with a
way to track orders. To show this, let us consider the product
\beq
N_n= {\cal N}_n {\cal N}_{n-1}\cdots {\cal N}_1  =
\left(p_n + x q_n + x^2 r_n \right)\cdots\left(p_1 + x q_1 + x^2 r_1 \right)
\enq

\vv \nin From the preceding development, making the substitutions $b_i \rightarrow
x b_i$ and  $c_i \rightarrow x^2 c_i$, we find

\beq N_n= x^{n-1} \left( P_n + x Q_n + x^2 R_n \right) \enq with
\beq P_n   = A_n \g(T+Z),~~~~Q_n = B_n,~~~~R_n = C_n \g(T-Z) \enq

\vv \nin  where the matrices $A_n$, $B_n$ and $C_n$ satisfy the
recurrence relations (\ref{REC1}). 

\vv \nin It is worth noticing that in $N_n$ the overall power factor $x^{n-1}$
fixes the lowest order (i.e. $n-1$) of the terms entering that matrix.
Moreover, terms of different orders are separated still in a clean way,
and hierarchy of orders is preserved : $A_n$ is of order $n-1$, $B_n$
is of order $n$ and $C_n$ is of order $n+1$. Notice also that in some
sense matrices $\g(T+Z)$ and $\g(T-Z)$ play the role of projectors
with respect to orders.

\vv \nin The same kind of treatment can be applied as well to a
matrix like
\beq
{\cal N}' = \g(\ep^{\star})[m+\g(Q')]
\enq

\nin and products of such matrices, so that the matrix ${\cal M}$ in
(\ref{AMB}) can be
written in the form

\begin{eqnarray}
&{\cal M} = \left[\g(T+Z)\bar{A'} + \bar{B'} + \g(T-Z)\bar{C'}\right]
\g(T-Z)\times & \nonumber \\
&\left[A\g(T+Z) + B + C\g(T-Z)\right]&
\end{eqnarray}

\nin i.e.

\begin{eqnarray}
&{\cal M} = \left[\g(T+Z)\bar{A'} + \bar{B'}\right]
\g(T-Z)\left[A\g(T+Z) + B \right]&
\end{eqnarray}
\begin{eqnarray}
&{\cal M} = 4 \bar{A'}^{(1)}\g(T+Z)A^{(1)}+ \bar{A'}^{(1)}\g(T+Z)
\g(T-Z)B + &
\label{AMP2} \nonumber \\
&\bar{B'}\g(T-Z)\g(T+Z)A^{(1)}+ \bar{B'}\g(T-Z)B &
\end{eqnarray}

\vv \nin It is interesting to notice that taking the light-like 4-vector $T-Z$ as
a polarization 4-vector of the virtual photon simply kills the higher
order tems contained in factors $\g(T-Z) C$. Thus, the matrix (\ref{AMP2})
comprises a first piece $\bar{A}\g(T+Z)A^{(1)}$ the order of which is
$p -2 + N- p -1 = n_{\g} -2$. This is the lowest order term of the matrix.
There are then two terms $\bar{A}\g(T+Z) \g(T-Z)B$ and
$\bar{B}\g(T-Z)\g(T+Z)A^{(1)}$ of order $n_{\g}-1$. The last term
$\bar{B}\g(T-Z)B$ is of order $n_{\g}$. We already know that the
amplitude $J$ in (\ref{AMB}) should be of order $n_{\g}$. So, we may expect
that when the matrix ${\cal M}$ is sandwiched in the scalar
product $J$, the lowest order term of
(\ref{AMP2}) picks up terms of overall order $2$ coming from the external
spinors, while the terms of order $n_{\g}-1$ and the last term of order
$n_{\g}$ pick up terms of overall order $1$ and $0$ respectively.
The least we can
say is that we should be very careful with respect to orders
when we compute the
amplitude in the framework of jet-like kinematics approximation, because
all terms conspire to produce an amplitude of final order $n_{\g}$.
Hopefully, as will be seen in the next section, to find out the
contribution due to the first term of (\ref{AMP2}), it is sufficient
to consider the development
of external spinors up to the order one only, because this term picks up
the terms of order one in each of the two external spinors !

\vv \nin To end up this section, let us make the following comments.

\vv \nin  a) Why did we decide to consider the effect of the product of
matrices $\left[m+\g(Q)\right]$ and $\g(\ep^{\star})$ instead of considering
the effect of each matrix separately ? The former matrix involves
terms of order zero, one and two with the same structure as in
(\ref{matrix2}). In contrast with this, the second matrix is a combination of
terms of order zero and one only, and has a different matrix structure :
the zero order term is $\g(E^{\star})$ and the first order term has
the factor $\g(T-Z)$. To obtain for this matrix a structure similar to
(\ref{matrix2}), it would have been necessary to multiply it by some first
order term, say the lepton mass. But, anyway, we know that second order
terms should not be discarded a priori. So instead of using some
complicated trick, we found it more convenient to
consider the product of these two matrices, which takes on the structure
(\ref{matrix2}) that leads to the nice algebra described above with a clean
separation of orders.

\vv \nin b) The introduction of the parameter $x$ in (\ref{matrix4})
has its origin in the observation that a boost of rapidity $\chi$ along
the $Z-$axis transforms $\g(T+Z)$, $\g(E^{\star})$ and $\g(T-Z)$ into,
respectively, $\exp(\chi) \g(T+Z)$, $\g(E^{\star})$ and
$\exp(-\chi) \g(T-Z)$, and a matrix like (\ref{matrix2}) is transformed into
$\exp(\chi) \left[a \g(T+Z) + \exp(-\chi) b + \exp(-2 \chi) c \g(T-Z)\right]$.
So, it appears that a boost along the $Z-$axis distinguishes between
the orders. This simple observation led us to the idea of introducing
the parameter $x$.

\subsection{Further possible treatment of the problem}

\vv \nin Taking into account the first relation in (\ref{spinor1}), it is convenient to
rewrite the spinors $U_1$ associated with the incoming lepton in the form

\begin{eqnarray}
&U^{\lambda}_1 = \sqrt{m} \left[ \exp(\di{\chi_1\over 2})\gamma(T+Z) + \exp(-
\di{\chi_1\over 2})\gamma(T-Z) \right]U^{\lambda}_0 /2~~~~~~~
{\rm or}& \label{SPI1} \nonumber \\
&U^{\lambda}_1  = \sqrt{\di{E_1\over 2}}\left[\gamma(T+Z) \alpha +
\gamma(T-Z) \beta\right] &
\end{eqnarray}

\nin where

\beq
\alpha = \sqrt{\di{{2m} \over{E_1}}} \exp(\di{\chi_1\over 2}) U^{\lambda}_0 =
\di{1\over 2}\left[~
\sqrt{1+m/E_1} + \sqrt{1-m/E_1}~\right] U^{\lambda}_0
\enq
and

\beq
\beta = \sqrt{\di{{2m} \over{E_1}}} \exp(-\di{\chi_1\over 2}) U^{\lambda}_0 =
\di{1\over 2}\left[~
\sqrt{1+m/E_1} - \sqrt{1-m/E_1}~\right] U^{\lambda}_0
\enq

\vv \nin Notice that in the jet-like kinematics approximation, $\beta$ is
found less than $\alpha$ by one order.

\vv \nin Now, let us see what is the result of the application of an elementary
matrix ${\cal N}$
on a spinor $\Psi = \gamma(T+Z) \alpha + \gamma(T-Z) \beta $. We get
\begin{eqnarray}
&{\cal N}\Psi = \left[ a \gamma(T+Z) + b + c \gamma(T-Z)\right]
\left[\gamma(T+Z) \alpha + \gamma(T-Z) \beta \right] = & \nonumber \\
& a \gamma(T+Z)\gamma(T-Z)\beta+ b \left[\gamma(T+Z) \alpha + \gamma(T-Z)
\beta \right] + & \\
&c \gamma(T-Z)\gamma(T+Z) \alpha & \nonumber
\end{eqnarray}

\nin But, from (\ref{fund2}), we have

\begin{eqnarray}
& a \gamma(T+Z)\gamma(T-Z)  = \gamma(T+Z) a^{(1)} \gamma(T-Z)  & \nonumber \\
&b \gamma(T-Z)  = \gamma(T-Z) b^{(1)},~~~b \gamma(T+Z)  =
\gamma(T+Z) b^{(2)} &  \\
&c \gamma(T-Z)\gamma(T+Z)=\gamma(T-Z)  c^{(1)} \gamma(T+Z)  & \nonumber
\end{eqnarray}

\nin Then,

\begin{eqnarray}
&{\cal N}\Psi = \gamma(T+Z)~a^{(1)}~\gamma(T-Z)~\beta +
\gamma(T+Z)~b^{(2)}~\alpha  + \gamma(T-Z)~b^{(1)}~\beta~+ & \nonumber \\
&\gamma(T-Z)~c^{(1)}~\gamma(T+Z)~\alpha~=~
\gamma(T+Z)~\left[ b^{(2)}~\alpha + a^{(1)}~\gamma(T-Z)~\beta
\right] + & \nonumber \\
& \gamma(T-Z)~
\left[ b^{(1)}~\beta + c^{(1)}~\gamma(T+Z) \alpha~ \right] &
\end{eqnarray}

\nin which is of the form

\beq
\Psi' = \gamma(T+Z) \alpha' + \gamma(T-Z) \beta'
\enq

\nin provided we set

\beq
 \alpha'  = b^{(2)} \alpha  + a^{(1)} \gamma(T-Z) \beta ~~~~~
{\rm and}~~~~~
\beta' = b^{(1)} \beta + c^{(1)} \gamma(T+Z) \alpha
\enq

\nin These two last relations may be conveniently recast in matrix form

\beq
\left( \begin{array}{c}
 \alpha'  \\
\beta'
\end{array} \right) = {\cal Y} \left( \begin{array}{c}
 \alpha  \\
\beta
\end{array} \right)
\enq

\nin with the 8X8 matrix ${\cal Y}$ given by

\beq
{\cal Y} = \left( \begin{array}{cc}
b^{(2)} & a^{(1)} \gamma(T-Z) \\
c^{(1)}\gamma(T+Z) & b^{(1)}
\end{array} \right)
\label{Y2}
\enq

\vv \nin Thus, the effect of the matrix ${\cal N}$ can be understood on a 8-dimensional space as follows

\begin{eqnarray}
&\Psi'  = {\cal N}\Psi \equiv \left( \begin{array}{cc}
\gamma(T+Z) & 0 \\
0 & \gamma(T-Z) \end{array} \right) {\cal Y}\left( \begin{array}{c}
 \alpha  \\
\beta
\end{array} \right)
\end{eqnarray}

\vv \nin It is then straightforward to find the effect of several matrices like
${\cal N}$ on $\Psi$

\begin{eqnarray}
&{\cal N}_p{\cal N}_{p-1}\cdots {\cal N}_1 \Psi \equiv
\left( \begin{array}{cc} \gamma(T+Z) & 0 \\
0 & \gamma(T-Z) \end{array} \right) {\cal Y}\left( \begin{array}{c}
 \alpha  \\
\beta
\end{array} \right)
\end{eqnarray}

\nin with

\beq
{\cal Y} = {\cal Y}_p {\cal Y}_{p-1} \cdots {\cal Y}_1
\enq

\vv \nin each matrix ${\cal Y}_k$ being given by an expression like (\ref{Y2}).

\vv \nin The main interest of such a presentation is that here again orders are
well separated. Thus in (76), the matrix elements at the right top are
of order zero ; the diagonal elements are of order one ; the matrix elements
at the left bottom are of second order. Acting on the vector of components
$\alpha$ and $\beta$, a matrix ${\cal Y}_i$ raises the order of these
components by one unit, but preserves the hierarchy of order between the
resulting components $\alpha'$ and $\beta'$ : $\beta'$ remains one order
less than $\alpha'$. Thus, this formalism provides a nice description of the
propagation of orders along the string of matrices in the amplitude
(\ref{AMB}).

\vv \nin Taking into account the results of the preceding section, the part of the
matrix (\ref{AMB}) that is at the right of $\g(T-Z)$ can then be
represented by
\beq
{\cal N}_r = \left( \begin{array}{cc} \gamma(T+Z) & 0 \\
0 & \gamma(T-Z) \end{array} \right){\cal Y}_{r}
\enq

\nin with

\beq
{\cal Y}_{r} = {\cal Y}_{p-1} {\cal Y}_{p-2} \cdots {\cal Y}_1 =
\left( \begin{array}{cc}
B^{(2)}_{r} & A^{(1)}_{r} \gamma(T-Z) \\
C^{(1)}_{r}\gamma(T+Z) & B^{(1)}_{r}
\end{array} \right)
\label{Y3}
\enq

\vv \nin Obviously, the same treatment can be applied as well to a
matrix like

\beq
\bar{{\cal N}} = \g(\ep^{\star})[m+\g(Q')]
\enq

\vv \nin and products of such matrices (notice the relation
$\bar{{\cal N}}_{\Lambda}(Q) = - \g_0 {\cal N}^{\dag}_{-\Lambda}(Q) \g_0$).
Thus, the part of the
matrix (\ref{AMB}) that is at the left of $\g(T-Z)$ can be
represented by
\beq
{\cal N}_{\ell} ={\cal Y}_{\ell} \left( \begin{array}{cc} \gamma(T+Z) & 0 \\
0 & \gamma(T-Z) \end{array} \right)
\enq

\vv \nin with

\beq
{\cal Y}_{\ell} = {\cal Y}_{N} {\cal Y}_{N-1} \cdots {\cal Y}_{p+1} =\left( \begin{array}{cc}
\bar{B}^{(2)}_{\ell} &\gamma(T+Z) \bar{C}^{(1)}_{\ell} \\
 \gamma(T-Z) \bar{A}^{(1)}_{\ell} & \bar{B}^{(1)}_{\ell}
\end{array} \right)
\label{Y4}
\enq

\vv \nin In appendix \ref{U3} we show that the spinors $U_3$ of the outgoing lepton can also be
written in a form analogous to the last
expression of $U_1$ in (\ref{SPI1}). Thus, we arrive at the following general
expression for the amplitude in (\ref{AMB}) (taking (\ref{AMP2}) into account)

\begin{eqnarray}
&J = \bar{U}_3 {\cal M} U_1 = \frac{1}{2}\sqrt{E_3 E_1}
\exp(i \lambda_3 \phi_3/2)
\left[\bar{\alpha}_3 \g(T+Z) + \bar{\beta}_3 \g(T-Z)\right] \times
 &\nonumber \\
&\left[4 \bar{A}^{(1)}_{\ell}\g(T+Z)A^{(1)}_r + \bar{A}^{(1)}_{\ell}\g(T+Z)
\g(T-Z)B_r +
\right. &  \\
&\left.\bar{B}_{\ell} \g(T-Z)\g(T+Z)A^{(1)}_r + \bar{B}_{\ell}
\g(T-Z)B_r \right]
\left[\g(T+Z)\alpha_1  +
 \g(T-Z)\beta_1 \right] &\nonumber
\end{eqnarray}

\nin i.e.

\begin{eqnarray}
&J = \frac{1}{2}\sqrt{E_3 E_1}  \exp(i \lambda_3 \phi_3/2)
\left[16~ \bar{\beta}_3 \bar{A}_{\ell} \g(T-Z) A_r \beta_1 + \right. &
\label{J1} \nonumber \\
& 4~ \bar{\beta}_3 \bar{A}_{\ell} \g(T-Z) B_r \g(T+Z)\alpha_1 +
4~ \bar{\alpha}_3 \g(T+Z)\bar{B}_{\ell} \g(T-Z) A_r \beta_1 + &  \\
&\left. 4~ \bar{\alpha}_3 \bar{B}^{(2)}_{\ell}\g(T+Z)B^{(2)}_r\alpha_1
\right] & \nonumber
\end{eqnarray}

\vv \nin From this last result we may draw the following important conclusions.
First, it is manifest that the expression in brackets in
(\ref{J1}) has lowest order $n_{\g}$. As expected, the lowest order term
in the matrix (\ref{AMP2}) selects the one order parts $\beta_3$ and
$\beta_1$ of external spinors to give a final expression of order $n_{\g}$.
In addition, the ratio $J/E_1$ is, as expected, independent on the incident
energy $E_1$.

\vv \nin Next, we arrive at a formidable nontrivial conclusion.
Up to now, we made exact calculations and (\ref{AMP2}) is an exact expression
for the generic form of impact factors. But if we go to jet-like kinematics
conditions, this expression shows us that to obtain the impact factor at
lowest order, it is sufficient to take in matrices ${\cal N}_i$ each term
equal to its lowest order expression. However, we should take each term into
account. In particular, we cannot generally discard second order terms such
as those proportionnal to $Q_{-}$ because they do participate to the
full amplitude.

\vv \nin Finally, let us remark that if the virtual photon is at the first
place in the Feynman diagram (reading the latter from right to left), we
should obviously take
$A_r = 0,~B_r = 1,~C_r =0$. If the virtual photon is at the last place,
we should take $\bar{A}_l = 0,~\bar{B}_l = 1,~\bar{C}_l =0$.

\subsection{Particular cases}

\vv \nin Simple general properties of impact factors may be established for some
particular \mbox{cases :} the case where all final particles are emitted in the
strict forward direction ; the case where one final photon takes away
the whole energy of the incoming lepton.
\vskip 0.25 cm

\noindent {\bf 1$^\circ$)} The emission in the strict forward direction : this has been already considered in \mbox{section 4.}  
\vv

\noindent {\bf 2$^\circ$)} One photon's $x$ is close to one.

\vv \nin Here, the $x$ of a photon of 4-momentum $k$ is the ratio $k_{+}/p_{1+}$,
which at lowest order is the fraction of energy taken away by the photon.

\vv \nin Assume first this energetic photon ($k_{+}/p_{1+} \rightarrow 1$)
being emitted at the $j$th place before the
vertex of the virtual photon. Then in every matrix ${\cal N}_i$
in (\ref{AMB}) with $ i < j$ we may take the 4-momenta of the
involved (soft) photons equal to zero, and we then get
\begin{eqnarray}
&{\cal N}_{j-1}{\cal N}_{j-2} \cdots {\cal N}_1 U^{\lambda}_1 =
 \left[m+\g(p_1)\right]\g(\epsilon^{\star}_{j-1}) \cdots
\left[m+\g(p_1)\right]\g(\epsilon^{\star}_1) U^{\lambda}_1 = &
\nonumber \\
& \left(2 p_1.\epsilon^{\star}_{j-1}\right)\cdots
\left(2 p_1.\epsilon^{\star}_1\right) U^{\lambda}_1
\end{eqnarray}

\nin Next, we turn to the effect of subsequent matrices. We said in the previous
section that it is possible to take each term equal to its lowest order
expression. Thus, in matrices ${\cal N}_m$ with $m \geq j$ we may
take $Q_{+} \equiv (p_1 - k_j)_{+} = 0$ (the 4-momenta of the other emitted
soft-photons are again set to zero). Thus, those matrices may be taken
in the form

\beq
{\cal N}_m \equiv \left[ m + \di{Q_{-}\over 2} \g(T-Z) + \g(Q_{T})\right]
\g(\epsilon^{\star}_{m})
\enq

\nin for propagators taking place before the vertex of the virtual photon, or

\beq
{\cal N}_n \equiv \g(\epsilon^{\star}_{n})\left[ m + \di{Q_{-}\over 2}
\g(T-Z) + \g(Q_{T})\right]
\enq

\nin for propagators involved after that vertex.

\vv \nin For both types of matrices the application of the matrix $\g(T-Z)$
kills the second order terms $\propto Q_{-}$'s and also the term
$\propto \xi$'s in matrices $\g(\epsilon^{\star})$. We thus get
\begin{eqnarray}
&\bar{{\cal N}}_N \cdots \bar{{\cal N}}_{p+1}
\g(T-Z) {\cal N}_{p-1} \cdots {\cal N}_j U^{\lambda}_1 \propto  &
\nonumber \\
& \g(E^{\star}_{N})\left[ m + \g(Q_{T N}) \right] \cdots
 \g(E^{\star}_{p+1})\left[m+ \g(Q_{T p+1}) \right] \g(T-Z) \times
& \nonumber \\
& \left[m+ \g(Q_{T p-1}) \right] \g(E^{\star}_{p-1}) \cdots
\left[ m + \g(Q_{T j}) \right]  \g(E^{\star}_{j}) U^{\lambda}_1  &
\end{eqnarray}

\nin But (see Eq. (\ref{TRANSH}) in appendix \ref{useformul}) 

\beq
 \g(E^{\star}_{j}) U^{\lambda}_1 = - 2 ~\lambda~ \sqrt{2}~\delta_{\Lambda_j,
2 \lambda}~ V^{-\lambda}_1
\enq

\vv \nin Because of the factor $\delta_{\Lambda_j, 2 \lambda}$, we conclude
that in such a case, the initial lepton ``transmits'' its
helicity to that $j$th energetic photon. We obtain the same conclusion
if the energetic photon is emitted after the vertex of the virtual photon.
Thus, this property holds true for the whole impact factor itself (where
all graphs are taken into account).


\subsection{Helicity properties of matrix factors
$A$, $B$ and $C$}

\vv \nin From the general relations (\ref{FUN1}-\ref{FUN3}), we obtain
\beq
\g(T+Z) \g(T-Z) A  = \g(T+Z) A^{(1)} \g(T-Z) =  A \g(T+Z) \g(T-Z)
\enq
thus, $A$ commutes with $\g(T+Z) \g(T-Z)$ or, equivalently, with the boost
operator $\g(Z) \g(T)/2$. We get the same conclusion for $B$ and $C$ since
\beq
\g(T-Z) \g(T+Z) B  = \g(T-Z) B^{(1)} \g(T+Z) =  B \g(T-Z) \g(T+Z)
\enq
\nin and
\beq
C \g(T-Z) \g(T+Z)   = \g(T-Z) C^{(1)} \g(T+Z) =   \g(T-Z) \g(T+Z) C
\enq

\vv \nin As the helicity operator is given by $S_Z =\g_5 \g(Z) \g(T)/2$, the helicity
properties of $A$, $B$ and $C$ are then determined by their commutation rule
with the chiral operator $\g_5$. From (\ref{matrix3}) it is seen that each of
the elementary matrices $a$, $b$ and $c$ can be divided in two parts :
a part proportional to the lepton mass $m$ and a part that does not contain
the lepton mass. Let us label the first one by a (o) for ``odd'' and the
second one by a (e) for ``even'' :
\beq
a^{o}  = 0,~~~~a^{e}  = -\di{Q_{+}\over 2} \g(E^{\star})
\enq
\beq
b^{o}  = m \g(E^{\star}),~~~~b^{e}  = \g(Q_T) \g(E^{\star}) +
\xi \di{Q_{+}\over 2} \g(T+Z) \g(T-Z)
\enq
\beq
c^{o}  = m \xi ,~~~~c^{e}  = \xi \g(Q_T) - \di{Q_{-}\over 2} \g(E^{\star})
\enq

\vv \nin Taking into account the fact that the matrices $\g(T\pm Z)$ are helicity
conserving, and that the ``transverse'' matrices $\g(E^{(\pm)})$ are helicity
flipping, it appears that

\biz
\item the odd parts $a^{o}$ and $c^{o}$ are helicity conserving while the
even parts
$a^{e}$ and $c^{e}$ do not conserve helicity ; this is also true for
$a^{(1)}$ and $c^{(1)}$ ;

\item the odd part $b^{o}$ does not conserve helicity whereas the even part
$b^{e}$ conserves helicity ; this is also true for $b^{(1)}$ and $b^{(2)}$.
\eiz

\vv \nin These properties can be generalized by recurrence to any matrix factors
$A_n$, $B_n$ and $C_n$. It is always possible to divide these factors
in a part involving odd powers of $m$ (labelled by a (o)) and a part
involving even powers of $m$ (labelled by a (e)) :
\begin{eqnarray}
&A_n = A^o_n + A^e_n & \nonumber \\
&B_n = B^o_n + B^e_n &  \\
&C_n = C^o_n + C^e_n & \nonumber
\end{eqnarray}

\vv \nin Let us assume the above itemized properties to be true for the rank $n$. Then,
from the recurrence relations (\ref{REC1}) we have

\begin{eqnarray}
&A^o_{n+1} = b^o_{n+1} A^e_n + b^e_{n+1} A^o_n + a^e_{n+1} B^{(1)o}_n
+ a^o_{n+1} B^{(1)e}_n & \nonumber \\
&A^e_{n+1} = b^o_{n+1} A^o_n + b^e_{n+1} A^e_n + a^o_{n+1} B^{(1)o}_n
+ a^e_{n+1} B^{(1)e}_n & \nonumber \\
&B^o_{n+1} = b^o_{n+1} B^e_n + b^e_{n+1} B^o_n + \g(T-Z) c^{(1)e}_{n+1}~
A^o_n \g(T+Z) + & \nonumber \\
& \g(T-Z) c^{(1)o}_{n+1}~A^e_n \g(T+Z) +\g(T+Z) a^{(1)o}_{n+1}~C^e_n \g(T+Z)
+    &  \nonumber \\
&  \g(T+Z) a^{(1)e}_{n+1}~C^o_n \g(T+Z) &  \\
&B^e_{n+1} = b^o_{n+1} B^o_n + b^e_{n+1} B^e_n + \g(T-Z) c^{(1)o}_{n+1}~
A^o_n \g(T+Z) + & \nonumber \\
& \g(T-Z) c^{(1)e}_{n+1}~A^e_n \g(T+Z) +\g(T+Z) a^{(1)o}_{n+1}~C^o_n \g(T+Z)
+    &  \nonumber \\
&  \g(T+Z) a^{(1)e}_{n+1}~C^e_n \g(T+Z) & \nonumber \\
&C^o_{n+1} = c^o_{n+1} B^{(2)e}_n + c^e_{n+1} B^{(2)o}_n +
b^o_{n+1} C^e_n + b^e_{n+1} C^o_n  & \nonumber \\
&C^e_{n+1} = c^o_{n+1} B^{(2)o}_n + c^e_{n+1} B^{(2)e}_n +
b^o_{n+1} C^o_n + b^e_{n+1} C^e_n  & \nonumber
\end{eqnarray}

\vv \nin From these decompositions it is not difficult to check
that the properties are also true for the rank $n+1$, QED.

\subsection{Helicity transitions in connexion with mass terms}

\vv \nin It is well known that in massless QED or QCD, transition amplitudes are
helicity conserving. This is due to the vector nature of the gauge particles,
photon or gluon. As is currently observed, helicity flips are due
to mass terms. The same kind of rule should be expected for the
presently studied impact factors. We would like to point out here that
\biz
\item amplitudes that conserve lepton helicity (HC amplitudes)
are even with respect to the lepton
mass (i.e. they do not change their sign when $m \rightarrow - m$) ;

\item amplitudes that do not conserve lepton helicity (HNC amplitudes) are
odd with respect to the lepton mass (i.e. they change their sign when
$m \rightarrow - m$).
\eiz

\vv \nin Let us examine from this point of view the structure of (\ref{J1}). Each
term in brackets may be again divided into a part (e) and a part (o).
For example, consider the first term. The spinor $\beta_1$ is odd and does
not involve a flip in helicity ($\theta_1 = 0$). We have
\begin{eqnarray}
&\bar{\beta}_3 \bar{A}_{\ell} \g(T-Z) A_r \beta_1 =\bar{\beta}^o_3
\bar{A}^e_{\ell} \g(T-Z) A^e_r \beta^o_1 +
 \bar{\beta}^o_3 \bar{A}^o_{\ell} \g(T-Z) A^o_r \beta^o_1 + & \nonumber \\
& \bar{\beta}^e_3 \bar{A}^e_{\ell} \g(T-Z) A^o_r \beta^o_1 +
\bar{\beta}^e_3 \bar{A}^o_{\ell} \g(T-Z) A^e_r \beta^o_1 + & \nonumber \\
&\bar{\beta}^e_3 \bar{A}^e_{\ell} \g(T-Z) A^e_r \beta^o_1 +
\bar{\beta}^e_3 \bar{A}^o_{\ell} \g(T-Z) A^o_r \beta^o_1 +
 &  \\
&\bar{\beta}^o_3 \bar{A}^e_{\ell} \g(T-Z) A^o_r \beta^o_1 +
\bar{\beta}^o_3 \bar{A}^o_{\ell} \g(T-Z) A^e_r \beta^o_1
 & \nonumber
\end{eqnarray}

\vv \nin From the helicity properties of matrix factors $A^e$ and $A^o$ on one hand,
and helicity properties of $\bar{\beta}^e_3$ and $\bar{\beta}^o_3$ on the
other hand (see appendix \ref{U3}), one can easily derive that the four first terms
in the above expansion are helicity conserving and are even functions of
the lepton mass ; the last four terms are odd functions of the lepton mass
and induce a flip in helicity. An analogous analysis can be carried out
in the same way for all terms in (\ref{J1}) ; whence the above stated
properties.

\section{Appendix}

\subsection{About the spinors $U_3$ of the outgoing lepton \label{U3} }

\setcounter{equation}{0}
\renewcommand{\theequation}{\mbox{A.}\arabic{equation}}

\vv \nin From (\ref{TR1}) and (\ref{TR2}) the spinors $U_3$ of the outgoing lepton
are given by
\begin{eqnarray}
&U^{\lambda_3}_3 = \sqrt{m}~ {\cal S}_3~  U^{\lambda_3}_0 & \nonumber \\
& {\rm with}~~~~{\cal S}_3 = {\cal R}_Z(\phi_3) {\cal R}_Y(\theta_3)
{\cal H}_Z(\chi_3)  &
\end{eqnarray}
(with normalisation ${\bar U}_3 U_3 = 2 m$).

\vv \nin First, we have
\beq
{\cal H}_Z(\chi_3)U^{\lambda}_0 =\frac{1}{2}\exp(\di{\chi_3\over 2})
\left[ \g(T+Z) + \exp(-\chi_3)\g(T-Z) \right]U^{\lambda}_0
\enq

\vv \nin Secondly, taking into account that $S_Y \g(T \pm Z) = \g(T \mp Z) S_Y$,
applying the rotation ${\cal R}_Y(\theta_3)$ to $\g(T\pm Z)$ yields
\beq
 {\cal R}_Y(\theta_3) \g(T\pm Z) =  \cos(\theta_3/2)\g(T \pm Z) -2 i
\sin(\theta_3/2)\g(T \mp Z) S_Y
\enq

\nin Then
\begin{eqnarray}
&{\cal R}_Z(\phi_3) {\cal R}_Y(\theta_3) \g(T\pm Z) =
\cos(\theta_3/2)\g(T \pm Z){\cal R}_Z(\phi_3)~ + & \nonumber \\
& - 2 i
\sin(\theta_3/2)\g(T \mp Z) S_Y {\cal R}_Z(-\phi_3)  &
\end{eqnarray}

\nin so that

\begin{eqnarray}
&U^{\lambda}_3 = \sqrt{m}~ {\cal S}_3~U^{\lambda}_0 = \sqrt{m}~
\frac{1}{2}\exp(\di{\chi_3\over 2}) \times
 \label{SPI2} & \nonumber \\
&\left\{ \cos(\theta_3/2) \exp(-i \lambda \phi_3) \left[\g(T+Z) +
\exp(-\chi_3)\g(T-Z)\right]\right. + &  \\
&\left. - 2 i \sin(\theta_3/2)\exp(i \lambda \phi_3)\left[\g(T-Z) +
\exp(-\chi_3)\g(T+Z)\right] S_Y \right\}~U^{\lambda}_0 & \nonumber
\end{eqnarray}

\nin Eq. (\ref{SPI2}) can be nicely rewritten in the form
\beq
U^{\lambda}_3 = \exp(-i \lambda \phi_3) \sqrt{\di{E_3\over 2}}
\left[ \g(T+Z) \alpha_3 + \g(T-Z) \beta_3 \right]
\enq

\nin with

\begin{eqnarray}
& \alpha_3 =  P(\di{m \over E_3})~
\left\{ \cos(\theta_3/2) - 2 i  \exp(2 i \lambda \phi_3)\sin(\theta_3/2)
\exp(-\chi_3) S_Y\right\}~U^{\lambda}_0
\label{a3b3}& \nonumber \\
&{\rm where}~~ P(\di{m \over E_3}) =
\frac{1}{2}\sqrt{\di{ 2 m \over E_3}}~\exp(\chi_3/2), ~~{\rm and} & \\
& \beta_3 = P(\di{m \over E_3})~
\left\{ \cos(\theta_3/2)\exp(-\chi_3) - 2 i  \exp(2 i \lambda \phi_3)
\sin(\theta_3/2) S_Y\right\}~U^{\lambda}_0
& \nonumber
\end{eqnarray}

\nin Remind that $ \cosh(\chi_3/2) = \sqrt{(E_3/m +1)/2}$, so that
\beq
P(\di{m \over E_3}) = \frac{1}{2} \left[ \sqrt{1 + m/E_3}+ \sqrt{1 - m/E_3}
\right]
\enq

\nin It is important to notice that $P(u)$ is an even function of $u = m/E_3$.
Therefore, its development in powers of $u$ contains even powers only
and when $u \ll 1$ we may write
\beq
P(u) \approx 1 + O(u^2)
\enq

\nin In the framework of jet-like kinematics approximation, this means that
since it is sufficient to keep only terms of first order in the development
of spinors, $P(\di{m \over E_3})$ can be safely taken equal to $1$. Then,
since we have $E_3 \gg m$ and
$\theta_3 \ll 1$, keeping only terms up to first order leads to
the following approximations

\begin{eqnarray}
& \alpha_3 \approx ~U^{\lambda}_0,~~~ \beta_3 \approx
\left[\di{ m \over {2 E_3}}  - i \xi_3 S_Y\right]~U^{\lambda}_0
& \nonumber \\
&{\rm where}~~~ \xi_3 =   \exp(2 i \lambda \phi_3) \theta_3 &
\end{eqnarray}

\nin Here again, we note that the spinor $\beta_3$ to which the matrix $\g(T-Z)$
applies is one order less than the spinor $\alpha_3$ to which the matrix
$\g(T+Z)$ applies : $\beta_3$ is of order $1$ while $\alpha_3$ is of order $0$.

\vv \nin It is also worth noticing the following general properties of spinors one can
derive from (\ref{a3b3}). Each of the two spinors $\alpha_3$ and $\beta_3$
is made up of two pieces that have different helicity properties and
different behaviors with respect to the lepton mass. Thus, taking into account
the relation
\beq
P(\di{m \over E_3})\exp(-\chi_3) = \frac{1}{2} \left[ \sqrt{1 + m/E_3}
- \sqrt{1 - m/E_3}\right]
\enq
we observe that
\biz
\item in $\alpha_3$, the piece that has the same helicity as that of
$U^{\lambda}_0$ is an even function of the lepton mass $m$, while the part
$ \propto - i S_Y U^{\lambda}_0 = \lambda U^{-\lambda}_0 $ that has
an opposite helicity is an odd function of $m$ ;
\item on the contrary, in $\beta_3$, the part with no helicity-flip is an
odd function of $m$, while the part with helicity-flip is an even function
of $m$.
\eiz

\newpage
\subsection {On ``mass-parity" of QED amplitudes \label{mass-parity} }

\setcounter{equation}{0}
\renewcommand{\theequation}{\mbox{B.}\arabic{equation}}

\vv \nin Any QED amplitude has one of the following two forms ; either 

\beq \bar{U}^{\lambda_\ell}_\ell  \, {\cal T}\, U^{\lambda_k}_k \enq 

\vv \nin for a subprocess like ${\rm lepton}_k \rightarrow X + {\rm lepton}_\ell$\, where both leptons are of the same species, or 

\beq \bar{U}^{\lambda_\ell}_\ell  \, {\cal T}\, W^{ \lambda_k}_k \enq 

\vv \nin for a subprocess where a lepton pair is produced, where in that case $U$ is the spinor of the lepton and $W$ the spinor associated with its antiparticle.  In both cases, 
${\cal T}$ is a 4X4 transition matrix which is a succession of products 
of lepton propagator and $\gamma$-matrices describing vertices.

\vv \nin In order to eliminate the mass terms from the various numerators of 
propagators entering into ${\cal T}$, we may apply Dirac equation as many times 
as necessary. In this way, the effective matrix ${\cal T}$ is found as a linear 
combination of products of $\gamma$-matrices, each of these products 
containing an {\em odd} number of $\gamma$-matrices, due to the vector 
nature of the lepton-photon coupling. Therefore, ${\cal T}$ may be expressed in the quite general form 

\beq
{\cal T} \sim \g(A) + \g(B) \gamma_5
\label{TD} \enq

\vv \nin where $A$ and $B$ are 4-vectors depending on 4-momenta of 
particles taking part in the subprocess, and on polarization 4-vectors of possible outgoing 
photons. Since it is always possible to make the choice\footnote{Which is different from that in Eq. (\ref{FW41}).} 

\beq W^\lambda \equiv V^\lambda = \g_5\, U^\lambda  \label{conjugate} \enq

\vv \nin one then sees that to study the ``mass-parity" property of more general amplitudes, it is sufficient to consider the elementary matrix elements  

\beq \bar{U}^{\lambda_\ell}_\ell  \, \g_\mu \, U^{\lambda_k}_k\,,~~\bar{U}^{\lambda_\ell}_\ell  \, \g_\mu \g_5 \, U^{\lambda_k}_k  \label{vertex1} \enq

\vv \nin In fact, it appears more convenient to consider instead matrix elements 

\beq \bar{U}^{\lambda_\ell}_\ell  \, \g_\mu \left( 1 \pm \g_5 \right) \, U^{\lambda_k}_k \label{vertex2} \enq

\vv \nin Using the definition (\ref{sp1}), we get 

$$ \left( 1 \pm \g_5 \right) \, U^{\lambda} = \exp{(\pm \lambda \chi)} \,  \left( 1 \pm \g_5 \right) \, {U'}^{\lambda} ~~~{\rm with} $$
\beq U^{' \lambda}= \sqrt{m} \left[\exp(-i \lambda \varphi)
\cos(\di{\theta\over 2})U^{\lambda}_0 + 2 \lambda \exp(i \lambda \varphi)
\sin(\di{\theta\over 2})U^{- \lambda}_0 \right] \enq

\vv \nin Then, 

\beq\bar{U}^{\lambda_\ell}_\ell  \, \g_\mu \left( 1 \pm \g_5 \right) \, U^{\lambda_k}_k = m\, \exp(\pm [ \lambda_\ell \chi_\ell + \lambda_k \chi_k]) \,{\cal V}_\mu (\theta_l, \varphi_\ell ; \theta_k, \varphi_k) \enq 

\vv \nin where 

\begin{eqnarray} & {\cal V}_\mu (\theta_l, \varphi_\ell ; \theta_k, \varphi_k) =  \left[\exp(i \lambda_\ell \varphi_\ell)
\cos(\di{\theta_\ell \over 2}) \bar{U}^{\lambda_\ell}_0 + 2 \lambda_\ell \exp(- i\lambda_\ell \varphi_\ell)
\sin(\di{\theta\over 2})\bar{U}^{-\lambda_\ell}_0 \right] \times & \nonumber \\
& \g_\mu \left( 1 \pm \g_5 \right) \left[\exp(- i\lambda_k \varphi_k)
\cos(\di{\theta_k \over 2})U^{\lambda_k}_0 + 2 \,\lambda_k  \exp(i\lambda_k \varphi_k)
\sin(\di{\theta_k\over 2})U^{- \lambda_k}_0 \right]& 
\end{eqnarray}

\nin clearly does not depend on the mass $m$. Hence, the whole dependence of amplitudes (\ref{vertex2}) on $m$ is entirely contained in the factors 

\beq {\cal P}^{(\pm)} (\ell, \lambda_\ell \,; k , \lambda_k) =  m\, \exp(\pm [ \lambda_\ell \chi_\ell + \lambda_k \chi_k])  \enq

\vv \nin Using 

$$ \exp( \pm \lambda \chi) = \di{1 \over \sqrt{2}} \, \sqrt{\di{E\over m}} \left[\, \sqrt{1 + \di{m\over E}} \pm 2 \lambda \sqrt{1 - \di{m\over E}} \,\, \right]$$

\vv \nin we find 

$$ {\cal P}^{(\pm)} (\ell, \lambda_\ell \,; k , \lambda_k) =  \di{1\over 2} \sqrt{E_\ell E_k} \,  \left[ \, \sqrt{1 + \di{m\over E_\ell }} \pm 2 \lambda_\ell \sqrt{1 - \di{m\over E_\ell }} ~\right] \times $$
\beq \left[ \,\sqrt{1 + \di{m\over E_k }} \pm 2 \lambda_k \sqrt{1 - \di{m\over E_k }}~ \right] \label{parity1} \enq

\vv \nin It is then clear that only expressions (\ref{parity1}) for which $\lambda_\ell = \lambda_k$ are even function of $m$, whereas those for which $\lambda_\ell = - \lambda_k$ are odd functions of $m$. It is worth noticing here that 4-momenta of particles entering the composition of $A$ or $B$ in (\ref{TD}) are even functions of $m$ (through energies $E = \sqrt{m^2 + {\vec{p}}^2}$). Therefore, we may draw the following conclusion.  

\vv \nin Let us consider one of the Feynman diagrams describing a given QED process where, to simplify, only one species of leptons is assumed to be involved. In the corresponding amplitude, any lepton $\ell_1$ (any anti-lepton $\bar{\ell}_1$) is connected, either to another lepton $\ell_2$ (another anti-lepton $\bar{\ell}_2$) through a sequence of lepton propagators, or to an anti-lepton $\bar{\ell}_3$ (lepton $\ell_3$) through, for example, subprocesses $\gamma^\star \leftrightarrow \ell + \bar{\ell}$ or $\g_1 + \g_2 \leftrightarrow \ell + \bar{\ell}$ with real or virtual photons. We may thus associate all leptons and anti-leptons taking part in the process in binomials like $(\ell_1, \ell_2)$, $(\bar{\ell}_1, \bar{\ell}_2)$ and $(\ell, \bar{\ell})$. The ``mass-parity" of the amplitude of the diagram is then found as follows. Let $N_e$ the number of binomials having particles with the same helicities, $N_o$ that of binomials where particles have opposite helicities. Then, the ``mass-parity" of the amplitude is equal to that of \mbox{$N_o$ :} even is $N_o$ is even, odd if $N_o$ is odd. We know that the all set of Feynman diagrams describing the process under consideration can be simply obtained from that particular diagram by interchanging lines of leptons or lines of anti-leptons, or interchanging between them in an appropriate way lepton lines with anti-lepton lines.  But it is clear that such operations lead to new amplitudes that possess the same ``mass-parity". We thus conclude that the full amplitude describing a process for given helicities of particles has the same ``mass-parity"  as that given by any of the underlying Feynman diagrams. This does not mean however that all sub-amplitudes are of the same order with regard to the mass $m$. For example, we may find $N_o =0$ for some diagram, $N_o = 2$ for another one. This implies that the amplitude of the latter is $\propto m^2$ compared to that of the first one.


\subsection{Some useful formulas \label{useformul} } 

\setcounter{equation}{0}
\renewcommand{\theequation}{\mbox{C.}\arabic{equation}}

\vv \nin In Ref {\bf [4]} useful relations involving Dirac spinors are given. It can be shown that
for any spinor we have 

\beq
\g_\mu ~U^{\lambda} =  t_\mu ~U^{\lambda} - 2 \lambda ~z_\mu
~V^{\lambda} + 2~ \lambda ~\sqrt{2}~ \epsilon^{(2 \lambda)}_{\mu}~ V^{-\lambda}
\enq

\nin where $t$, $z$ and circular polarisations $\epsilon^{(2 \lambda)}$ with $\lambda = \pm 1/2$ are defined by eqs (\ref{hframe1}) and (\ref{hframe2}).  Hence,

\beq
\g(\epsilon^{(\Lambda)\star}) ~U^{\lambda} = -  2~ \lambda ~\sqrt{2} ~
\delta_{\Lambda, 2\lambda} ~ V^{-\lambda}
\enq

\nin In particular, for the ingoing lepton we may set $\epsilon^{(\Lambda)} = E^{(\Lambda)}$ (see (\ref{hframe1})) and 

\beq
\g(E^{(\Lambda)\star}) ~U^{\lambda}_1 = -  2~ \lambda ~\sqrt{2} ~
\delta_{\Lambda, 2\lambda} ~ V^{-\lambda}_1
\label{TRANSH} \enq

\vv \nin Defining, for any 4-vector $Q$

\beq
Q^{(\Lambda)}_T = Q_X + i \Lambda Q_Y = \Lambda \sqrt{2}E^{(\Lambda)}\cdot Q
\enq

\nin we also get

\beq
\g(Q_T) ~U^{\lambda}_0 = Q^{(2 \lambda)}_T~ V^{-\lambda}_0
\enq

\beq
\g(Q_T)~\g(E^{(\Lambda)\star}) ~U^{\lambda}_0 = 2~ \lambda ~\sqrt{2}
~\delta_{\Lambda, 2\lambda} ~ Q^{(- 2 \lambda)}_T~ U^{\lambda}_0
\enq

\nin We have

\beq
\g(T-Z )~U^{\lambda}_0 = U^{\lambda}_0 - 2~ \lambda ~V^{\lambda}_0
\enq

\nin and

\beq
\g(E^{(\Lambda)\star}) ~\g(T-Z )~U^{\lambda}_0 =- \sqrt{2}~\delta_{\Lambda,
2\lambda}~ \left[~U^{-\lambda}_0 + 2~ \lambda ~V^{-\lambda}_0 ~\right]
\enq

\beq
\g(E^{(\Lambda)\star}) ~\g(T-Z )~V^{\lambda}_0 =- \sqrt{2}~\delta_{\Lambda,
2\lambda}~ \left[~V^{-\lambda}_0 + 2~ \lambda ~U^{-\lambda}_0 ~\right]
\enq

\vv \nin In (\ref{Y2}) we have the matrix $a^{(1)} \g(T-Z) = Q_{+}\g(E^{(\Lambda)\star})
~\g(T-Z)/2$. Its action on the basis of spinors $U^{\lambda}_0,~
V^{\lambda}_0$ may be summed up by the following matrices :

\beq
a^{(1)} \g(T-Z) \equiv \di{Q_{+}\over 2}~\sqrt{2}~\left( \begin{array}{cccc}
0 & 0 & 0 & 0 \\
-1 & 0 & -1 & 0 \\
0 & 0 & 0 & 0 \\
-1 & 0 & -1 & 0 \\
\end{array} \right)~~~{\rm for}~~~\Lambda = + 1
\enq

\vv

\beq
a^{(1)} \g(T-Z) \equiv \di{Q_{+}\over 2}~\sqrt{2}~\left( \begin{array}{cccc}
0 & -1 & 0 & 1 \\
0 & 0 & 0 & 0 \\
0 & 1 & 0 & -1 \\
0 & 0 & 0 & 0 \\
\end{array} \right)~~~{\rm for}~~~\Lambda = - 1
\enq

\nin where lines as well as columns are arranged according to
the sequence
$U^{\uparrow}_0,~U^{\downarrow}_0,~V^{\uparrow}_0,~V^{\downarrow}_0 $.

\nin We also have
\beq
\g(Q_T)~\g(E^{(\Lambda)\star}) \equiv Q^{(-)}_T~\sqrt{2}~
\left( \begin{array}{cccc}
1 & 0 & 0 & 0 \\
0 & 0 & 0 & 0 \\
0 & 0 & 1 & 0 \\
0 & 0 & 0 & 0 \\
\end{array} \right)~~~{\rm for}~~~\Lambda = + 1
\enq
\beq
\g(Q_T)~\g(E^{(\Lambda)\star}) \equiv Q^{(+)}_T~\sqrt{2}~
\left( \begin{array}{cccc}
0 & 0 & 0 & 0 \\
0 & -1 & 0 & 0 \\
0 & 0 & 0 & 0 \\
0 & 0 & 0 & -1 \\
\end{array} \right)~~~{\rm for}~~~\Lambda = - 1
\enq

\nin Thus, the matrix $b^{(1)}$ in (\ref{Y2}) may be represented in the form :
\beq
b^{(1)} \equiv \sqrt{2}~
\left( \begin{array}{cccc}
Q^{(-)}_T & 0 & 0 & 0 \\
0 & 0 & - m & 0 \\
0 & 0 & Q^{(-)}_T & 0 \\
m & 0 & 0 & 0 \\
\end{array} \right)~~~{\rm for}~~~\Lambda = + 1
\enq

\beq
b^{(1)} \equiv -~\sqrt{2}~
\left( \begin{array}{cccc}
0 & 0 & 0 & - m \\
0 & Q^{(+)}_T & 0 & 0 \\
0 & m & 0 & 0 \\
0 & 0 & 0 &  Q^{(+)}_T \\
\end{array} \right)~~~{\rm for}~~~\Lambda = - 1
\enq

\nin We have
\beq
\g(T-Z)~\g(T+Z) = 2\left( 1 - \g(Z)\g(T)\right) = 2 \left( 1 -
2 \g_5 S_Z\right)
\enq

\nin so that
\beq
\g(T-Z)~\g(T+Z)~U^{\lambda}_0 = 2 \left( U^{\lambda}_0  - 2 \lambda
V^{\lambda}_0 \right)
\enq

\nin Thus
\beq
\g(T-Z)~\g(T+Z) \equiv ~2~
\left( \begin{array}{cccc}
1 & 0 & -1 & 0 \\
0 & 1 & 0 & 1 \\
-1 & 0 & 1 & 0 \\
0 & 1 & 0 & 1 \\
\end{array} \right)
\enq

\nin Let us define

\beq
\eta^{(\mp)}=\tan(\di{\theta\over 2})\exp(\mp i \phi)~~~{\rm and}~~~
   \zeta^{(\mp)} =  \di{Q_{+}\over 2}~\eta^{(\mp)}
\enq

\nin then, the matrix $b^{(2)}$ in (\ref{Y2}) has the form :

\beq
b^{(2)} \equiv \sqrt{2}~
\left( \begin{array}{cccc}
Q^{(-)}_T -\zeta^{(-)} \ & 0 & \zeta^{(-)} & 0 \\
0 & -\zeta^{(-)} & - m &  -\zeta^{(-)}  \\
\zeta^{(-)} & 0 & Q^{(-)}_T -\zeta^{(-)} & 0 \\
m & -\zeta^{(-)} & 0 &  -\zeta^{(-)}  \\
\end{array} \right)~~~{\rm for}~~~\Lambda = + 1
\enq

\beq
b^{(2)} \equiv -~\sqrt{2}~
\left( \begin{array}{cccc}
 -\zeta^{(+)} & 0 & \zeta^{(+)} & - m \\
0 & Q^{(+)}_T -\zeta^{(+)} & 0 &  -\zeta^{(+)} \\
\zeta^{(+)} & m &  -\zeta^{(+)} & 0 \\
0 &  -\zeta^{(+)} & 0 &  Q^{(+)}_T -\zeta^{(+)} \\
\end{array} \right)~~~{\rm for}~~~\Lambda = - 1
\enq

\nin We have

\beq
\g(T+Z )~U^{\lambda}_0 = U^{\lambda}_0 + 2~ \lambda ~V^{\lambda}_0
\enq
\beq
\g(T+Z )~V^{\lambda}_0 = -V^{\lambda}_0 - 2~ \lambda ~U^{\lambda}_0
\enq

\nin and

\beq
\g(E^{(\Lambda)\star}) ~\g(T+Z )~U^{\lambda}_0 =\sqrt{2}~\delta_{\Lambda,
2\lambda}~ \left[~U^{-\lambda}_0 - 2~ \lambda ~V^{-\lambda}_0 ~\right]
\enq
\beq
\g(E^{(\Lambda)\star}) ~\g(T+Z )~V^{\lambda}_0 =\sqrt{2}~\delta_{\Lambda,
2\lambda}~ \left[~V^{-\lambda}_0 - 2~ \lambda ~U^{-\lambda}_0 ~\right]
\enq
\beq
\g(Q_T) ~\g(T+Z )~U^{\lambda}_0 = Q^{(2\lambda)}_T~ \left[~
V^{-\lambda}_0 - 2~ \lambda ~U^{-\lambda}_0 ~\right]
\enq
\beq
\g(Q_T) ~\g(T+Z )~V^{\lambda}_0 = Q^{(2\lambda)}_T ~ \left[~
U^{-\lambda}_0 - 2~ \lambda ~V^{-\lambda}_0 ~\right]
\enq

\vv \nin then, the matrix $c^{(1)} \g(T+Z)$ in (\ref{Y2}) has the form :

\begin{eqnarray}
c^{(1)}~\g(T+Z) \equiv \di{\sqrt{2}\over 2}~ \eta^{(-)}~
\left( \begin{array}{cccc}
-m   & Q^{(-)}_T  & m  & Q^{(-)}_T  \\
- Q^{(+)}_T & - m & Q^{(+)}_T & -m   \\
-m & Q^{(-)}_T & m & Q^{(-)}_T  \\
 Q^{(+)}_T &  m  &- Q^{(+)}_T & m    \\
\end{array} \right) + \nonumber \\
\di{\sqrt{2}\over 2}~ Q_{-}\left( \begin{array}{cccc}
0 & 0 & 0 & 0  \\
1 & 0 & -1 & 0   \\
0 & 0 & 0 & 0  \\
-1 & 0 & 1 & 0    \\
\end{array} \right)~~~{\rm for}~~~\Lambda = + 1
\end{eqnarray}

\nin and

\begin{eqnarray}
c^{(1)}~\g(T+Z) \equiv \di{\sqrt{2}\over 2}~ \eta^{(+)}~
\left( \begin{array}{cccc}
m  & - Q^{(-)}_T  & - m  & - Q^{(-)}_T  \\
Q^{(+)}_T & m & - Q^{(+)}_T & m   \\
m & - Q^{(-)}_T & -m & -Q^{(-)}_T  \\
-Q^{(+)}_T & - m  & Q^{(+)}_T & - m    \\
\end{array} \right) + \nonumber \\
\di{\sqrt{2}\over 2}~ Q_{-}\left( \begin{array}{cccc}
0 & 1  & 0 & 1  \\
0 & 0  & 0 & 0   \\
0 & 1 & 0 & 1  \\
0 & 0  & 0 & 0    \\
\end{array} \right)~~~{\rm for}~~~\Lambda = - 1
\end{eqnarray}

\nin From these relations we derive the following expressions of the 8X8 matrices
${\cal Y}_r$ and ${\cal Y}_{\l}$

\begin{eqnarray}
&{\cal Y}_r \equiv \di{\sqrt{2}\over 2} \times & \nonumber \\
 & &\nonumber \\
& \left( \begin{array}{cccccccc}
2 Q^{(-)}_T -2\zeta^{(-)} & 0 &2\zeta^{(-)} & 0 &0 &0 &0 &0  \\
 0  &- 2\zeta^{(-)} & - 2m & -2\zeta^{(-)} &-Q_{+}&0&- Q_{+}& 0 \\
2\zeta^{(-)} & 0 & 2 Q^{(-)}_T -2\zeta^{(-)} &0&0&0&0&0 \\
 2m &- 2\zeta^{(-)} & 0 &  -2\zeta^{(-)} &-Q_{+}&0&-Q_{+}& 0  \\
-m \eta^{(-)} & \eta^{(-)}Q^{(-)}_T &m \eta^{(-)} &\eta^{(-)}Q^{(-)}_T &
2Q^{(-)}_T &0 &0 &0 \\
 Q_{-}-\eta^{(-)} Q^{(+)}_T &-m \eta^{(-)} &\eta^{(-)} Q^{(+)}_T- Q_{-} &
-m \eta^{(-)} &0 &0 &-2m &0 \\
-m \eta^{(-)} &\eta^{(-)}Q^{(-)}_T &m \eta^{(-)} &\eta^{(-)} Q^{(-)}_T &0 &0 &
2Q^{(-)}_T &0 \\
\eta^{(-)} Q^{(+)}_T- Q_{-} &m \eta^{(-)} & Q_{-}-\eta^{(-)} Q^{(+)}_T &
m \eta^{(-)} &2m &0 &0 &0 \\
\end{array} \right)& \nonumber \\
 &  &\nonumber \\
& &\nonumber \\
&{\rm for}~~~~\Lambda = + 1 &
\end{eqnarray}

\begin{eqnarray}
&{\cal Y}_r \equiv   \di{\sqrt{2}\over 2} \times & \nonumber \\
 & &\nonumber \\
& \left( \begin{array}{cccccccc}
2\zeta^{(+)}& 0 &-2\zeta^{(+)}& 2 m & 0 & -Q_{+} & 0 & Q_{+}  \\
0 & 2(\zeta^{(+)}-Q^{(+)}_T) & 0 & 2\zeta^{(+)} & 0 & 0 & 0 & 0 \\
-2\zeta^{(+)} & -2m &2\zeta^{(+)} & 0 & 0 &Q_{+}& 0 &-Q_{+} \\
0 & 2\zeta^{(+)} & 0 & 2(\zeta^{(+)}- Q^{(+)}_T) & 0 & 0 & 0 & 0  \\
m \eta^{(+)} &Q_{-}- \eta^{(+)}Q^{(-)}_T &- m \eta^{(+)} &Q_{-}-\eta^{(+)}
Q^{(-)}_T & 0 & 0 & 0 & 2m \\
\eta^{(+)} Q^{(+)}_T &m \eta^{(+)} &-\eta^{(+)} Q^{(+)}_T &
m \eta^{(+)} & 0 & -2 Q^{(+)}_T & 0 & 0 \\
m \eta^{(+)} &Q_{-}-\eta^{(+)} Q^{(-)}_T  & -m \eta^{(+)} &
Q_{-}-\eta^{(+)} Q^{(-)}_T & 0 & -2m & 0 & 0 \\
-\eta^{(+)} Q^{(+)}_T &- m \eta^{(+)} &\eta^{(+)} Q^{(+)}_T &
-m \eta^{(+)} & 0 & 0 & 0 & -2 Q^{(+)}_T \\
\end{array} \right)& \nonumber \\
 &  &\nonumber \\
& &\nonumber \\
&{\rm for}~~~~\Lambda = - 1 &
\end{eqnarray}

\begin{eqnarray}
{\cal Y}_{\l}^{\Lambda} =- \Gamma_0
\left[{\cal Y}_{r}^{-\Lambda}\right]^{\dag} \Gamma_0  &{\rm with} &
 \Gamma_0 = \left( \begin{array}{cc}
\g_0 & 0   \\
0  &\g_0 \\
\end{array} \right)
\end{eqnarray}
\nopagebreak
\vskip 2 true cm

\newpage

\centerline { REFERENCES}\par
\begin{description}

\item [{1}] C. Carimalo, Thesis, 1977, Paris (unpublished).

\item [{2}]  E.A. Kuraev, A. Schiller, V.G. Serbo, D.V. Serebryakova, Eur. Phys. J. C4 (1998) 631-639.  \par

\item [{3}] C. Carimalo, A. Schiller, V.G. Serbo, Eur.Phys.J. C23 (2002) 633-649 ; Eur.Phys.J. C29 (2003) 341-351. 

\item [{4}]  C. Carimalo, J.Math.Phys. 34 (1993) 4930-4963.

\end{description}

\end{document}